\documentclass[journal, one column]{IEEEtran}
\usepackage{amsmath,amsfonts}
\usepackage{algorithm2e}
\usepackage{array}
\usepackage[caption=false,font=normalsize,labelfont=sf,textfont=sf]{subfig}
\usepackage{textcomp}
\usepackage{stfloats}
\usepackage{url}
\usepackage{verbatim}
\usepackage{graphicx}
\usepackage{cite}
\usepackage{subfig} 
\usepackage{amsthm}

\usepackage[margin=0.65in, bottom=1in]{geometry}
\usepackage{makecell}
\usepackage{multirow} 

\usepackage{amsthm}

\usepackage{graphicx} 
\usepackage{amsmath}
\usepackage{amsmath}
\usepackage{amsmath, amssymb}
\usepackage{xcolor}	
\usepackage{amsmath} 
\allowdisplaybreaks[4]

\usepackage{algpseudocode}
\usepackage{graphics}
\usepackage{amsmath, amsthm}

\theoremstyle{plain}

\begin{document}


    \title{Inner and Outer Bounds on the Secrecy Capacity of Degraded Broadcast Channels with RMSI and Transmitter CSI
}

	\author{IEEE Publication Technology,~\IEEEmembership{Staff,~IEEE,}
	}

\author{Saeid Pakravan, Mohsen Ahmadzadeh, and Ghosheh Abed Hodtani

\thanks{
S. Pakravan is with the Department of Computer Science, University of Quebec in Montreal (UQAM), Montreal, QC, Canada.
email: pakravan.saeid@uqam.ca.} 
\thanks{
M.  Ahmadzadeh and G. Abed Hodtani are with the Department of Electric and Computer Engineering, Ferdowsi University, Mashhad, Iran. email: m.ahmadzadehbolghan@mail.um.ac.ir; hodtani@um.ac.ir.
}

}

	\maketitle

\begin{abstract}


This paper studies the secrecy capacity of a class of degraded broadcast channels in the presence of an external eavesdropper, where a transmitter aims to deliver two independent confidential messages to two legitimate receivers. The transmitter is assumed to have non-causal access to the channel state information (CSI), and each legitimate receiver possesses prior knowledge of the other receiver’s message, referred to as receiver message side information (RMSI). We consider two distinct scenarios: complementary RMSI, where each receiver knows only the other’s message, and non-complementary RMSI, where the SI does not perfectly align. For both scenarios, we derive novel inner bounds on the achievable secrecy rate region and present tight outer bounds, establishing the secrecy capacity region for the considered degraded channel settings. Unlike prior works, which primarily address general broadcast settings without secrecy constraints or omit key interactions between RMSI and CSI, our results provide a complete characterization of the secure communication limits under these conditions. Moreover, we extend our analysis to the Gaussian degraded broadcast channel, highlighting the pivotal role of CSI in enhancing secure transmission performance. Our findings demonstrate that the combination of RMSI and CSI can be strategically leveraged to expand the secrecy capacity region, thus offering new insights into secure multiuser communication system design.
\\

\end{abstract}

\section{Introduction}

Broadcast communication constitutes a cornerstone of contemporary wireless network architectures, underpinning a wide range of critical applications, including cellular downlink transmissions, satellite communications, and digital television broadcasting. The theoretical abstraction of this communication model is encapsulated by the broadcast channel (BC), a concept initially introduced by Cover in [1]. In this model, a single transmitter seeks to deliver distinct, independent messages to multiple receivers that operate without cooperation. Despite extensive research spanning several decades, characterizing the full capacity region of the general discrete memoryless broadcast channel remains one of the fundamental open problems in information theory.

Nevertheless, significant progress has been made in analyzing and deriving capacity results for specific subclasses of BCs. For example, degraded, more capable, less noisy, deterministic, and semi-deterministic BCs have been thoroughly investigated, each offering unique insights and tractable capacity characterizations. Marton’s pioneering work [2], along with subsequent contributions by Nair and El Gamal [3], established the best-known inner bounds on the capacity region of the general two-receiver BC. In the case of degraded BCs, Bergmans introduced the technique of superposition coding [4], which was later shown to be capacity-achieving through the converse results of Ahlswede and Körner [5], [6]. Further foundational contributions, such as those by Körner and Marton [7], have characterized the capacity regions for less noisy and more capable BCs, where superposition coding continues to serve as an optimal strategy [8], [9].

An important extension of the classical BC problem involves the integration of side information (SI), which refers to the availability of additional knowledge pertaining to the communication environment or the transmitted data [10-12]. The foundational concept of incorporating SI into communication models was first introduced by Shannon, who derived the capacity of point-to-point channels with causal SI at the transmitter [13]. This initial work paved the way for subsequent generalizations. Notably, Gel’fand and Pinsker extended the analysis to channels with non-causal SI at the transmitter [14], revealing how pre-available channel knowledge could be exploited to enhance communication rates. Further developments considered the presence of SI at both the transmitter and the receiver [15], leading to broader applicability in practical communication scenarios. These theoretical insights were later specialized to Gaussian channels [16], [17] and systems with partial channel state information (CSI) [18], which are more representative of real-world wireless environments where perfect CSI is rarely available.

Building upon these theoretical underpinnings, the study of broadcast channels with SI at the transmitter has gained considerable traction due to its practical relevance in wireless network deployments, particularly at base stations (BSs) equipped with partial or full environmental knowledge. Seminal work by Steinberg and Shamai [19], [20] extended Marton’s coding scheme to state-dependent BCs, deriving inner and outer bounds for scenarios with SI at the transmitter. Their contributions laid the groundwork for more generalized multiuser frameworks. For instance, Farsani and Marvasti advanced this line of research by analyzing the performance limits of multiuser BCs with transmitter-side SI in [21].

In parallel, the pursuit of secure communication over BCs has emerged as a vital research direction, especially in adversarial environments where information privacy must be preserved. Wyner's groundbreaking model of the wiretap channel [22] demonstrated that it is possible to achieve information-theoretic security without relying on pre-shared keys. This concept was later extended by Csiszár and Körner [23] to BCs with confidential messages, establishing the secrecy capacity region under a variety of assumptions. This foundational work spurred a series of studies exploring diverse secure BC configurations. These include the design of coding strategies to ensure that private messages remain confidential from unintended receivers [24], [25], the formulation of secrecy constraints under multi-eavesdropper scenarios [26], and the treatment of communication models with both common and private messages [27], [28]. These works have also incorporated message side information (MSI) at legitimate receivers, demonstrating improvements in secrecy rate regions.

In certain communication models, a portion of the transmitter's messages may already be available to the receivers as a priori information, referred to as receiver MSI (RMSI). The presence of RMSI significantly impacts the achievable secrecy rates by reducing the decoding burden at the receivers. While the secrecy capacity of BCs with RMSI has been explored in several studies [29-35], the interplay between RMSI and channel state information (CSI) at the transmitter remains underexplored.
While the interplay of CSI and RMSI has been explored
in prior literature, such as the unified inner bounds for BCs
presented in [36], [37], most of these
works focus on general capacity characterizations without
addressing secrecy constraints. In contrast, the secrecy setting
introduces new challenges in protecting confidential messages
from eavesdroppers, particularly when legitimate users possess
MSI and the transmitter exploits CSI.

This paper addresses these gaps by investigating the secrecy capacity region of the degraded BC with an external eavesdropper, non-causal CSI at the transmitter, and RMSI at the receivers. Unlike prior works that primarily focused on capacity regions without secrecy constraints or lacked treatment of RMSI structures, our study develops new inner and outer bounds for the secrecy capacity. Specifically, we focus on two classes of BCs: those with complementary RMSI and those with non-complementary RMSI.
Our motivation stems from the practical relevance of bidirectional relaying in three-node networks, where a relay facilitates communication between two nodes in two phases. In the multiple access channel (MAC) phase, both nodes transmit their messages to the relay, which decodes them. Assuming that the eavesdropper has no access to the messages during this phase, the subsequent broadcast phase aligns with the BC model studied in this work, featuring RMSI at legitimate receivers and CSI at the transmitter. Notably, the eavesdropper is assumed to have no access to the MSI.

For Gaussian BCs, the availability of CSI at the transmitter has been shown to enhance the achievable secrecy rate regions. Our study extends these results by demonstrating the additional benefits of RMSI in improving secrecy capacity.

The main contributions of this paper are as follows:

\begin{itemize}

\item We derive novel inner and outer bounds for the secrecy capacity region of BCs with CSI at the transmitter and RMSI at the receivers, providing theoretical insights into the impact of these features on secure communication.

\item We provide an in-depth comparison between complementary and non-complementary RMSI cases, highlighting the differing impacts of SI structure on secrecy performance.

\item We extend our analysis to the Gaussian BC, demonstrating how CSI and RMSI jointly contribute to enhancing secrecy and how the derived bounds apply in continuous-alphabet settings.

\end{itemize}

Our findings reveal that the inner bounds significantly outperform baseline schemes and that the availability of SI at receivers can be leveraged to improve secrecy, even when the eavesdropper has a statistically better channel. This work establishes the importance of combining CSI and RMSI in designing secure broadcast protocols and differentiates itself by its specific focus on secrecy under structured SI, a scenario not previously analyzed in depth.

The remainder of this paper is organized as follows. Section II introduces the system model and key definitions. In Section III, we present our main results, including achievable secrecy rate regions and outer bounds for both complementary and non-complementary RMSI scenarios. Section IV focuses on the Gaussian BC case, providing in-depth analysis and discussion. Finally, Section V concludes the paper with a summary of key findings.

\section{Channel Models and Definitions}

\subsection{Channel Models}

\begin{figure}[ht]
    \centering
 \includegraphics[
        width=0.8\textwidth,
        height=0.25\textheight
    ]{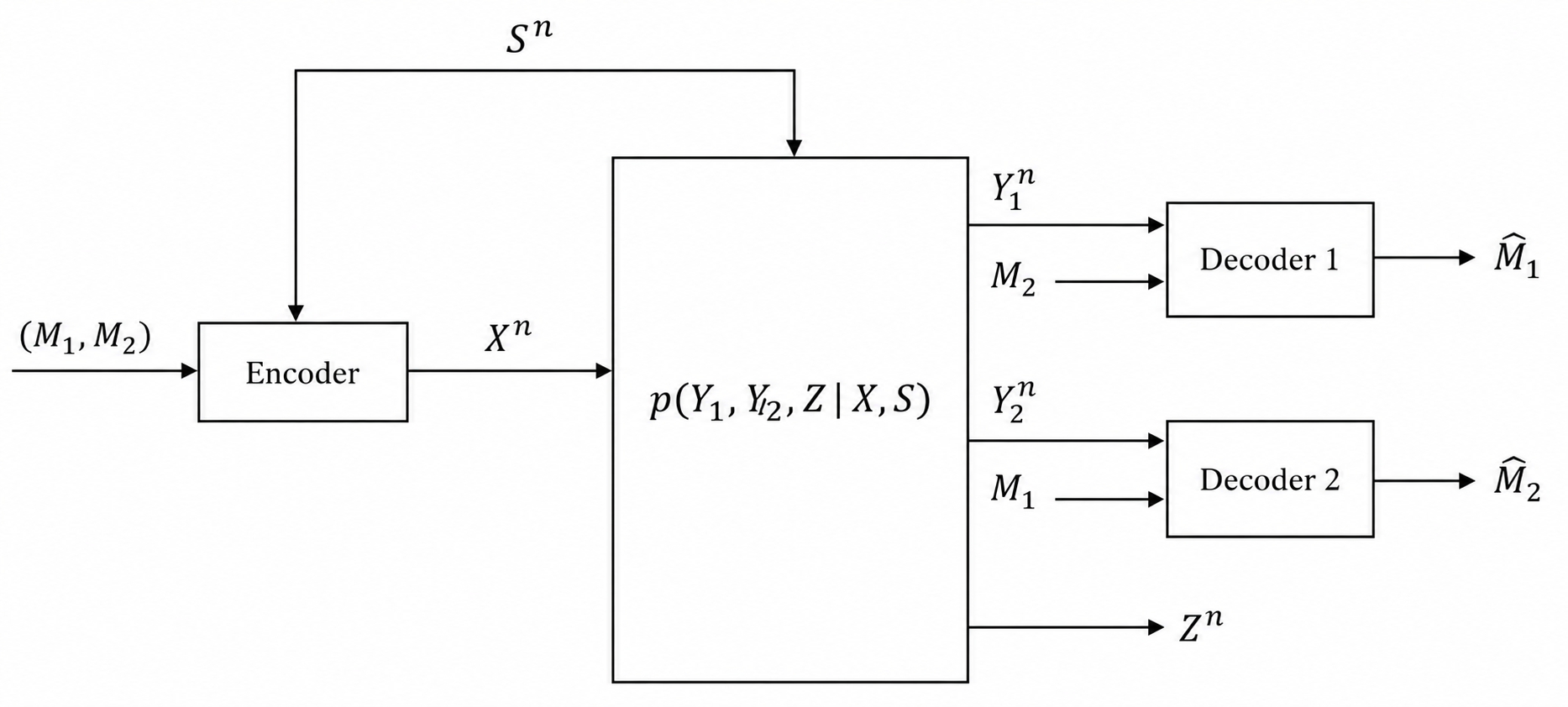}
    \caption{System model of a two-receiver BC with complementary RMSI, non-causal CSI at the transmitter, and an external eavesdropper.}
    \label{fig:fig1}
\end{figure}

\textbf{Model 1: Two-Receiver BC with Complementary RMSI.}  
The channel model for a two-receiver BC with complementary RMSI, non-causal CSI at the transmitter, and the presence of an eavesdropper is illustrated in Fig.~\ref{fig:fig1}. In this scenario, the two legitimate receivers are characterized by having complementary MSI, meaning the messages available to one receiver are not available to the other, ensuring distinct decoding requirements. The transmitter has access to non-causal CSI, which enables it to anticipate and adapt its transmission strategy to the state of the channel, enhancing the secrecy capacity against the eavesdropper.

\begin{figure}[ht]
    \centering
    \includegraphics[
        width=0.8\textwidth,
        height=0.25\textheight
    ]{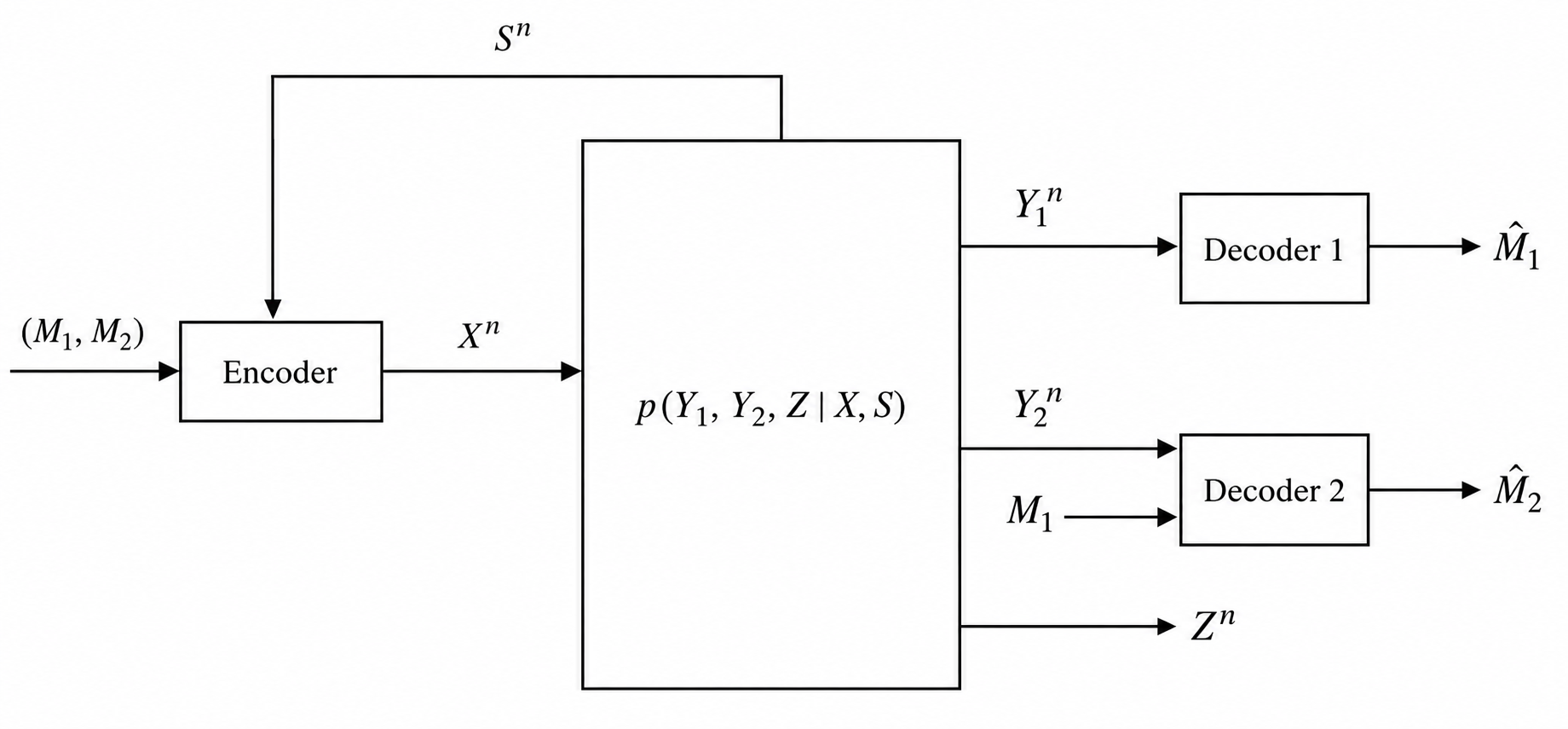}
    \caption{System model of a two-receiver BC with non-complementary RMSI, non-causal CSI at the transmitter, and an external eavesdropper.}
    \label{fig:fig2}
\end{figure}

\textbf{Model 2: Two-Receiver BC with Non-Complementary RMSI.}  
The channel model for the two-receiver BC with non-causal CSI at the transmitter and non-complementary RMSI in the presence of an eavesdropper is depicted in Fig.~\ref{fig:fig2}. The model retains the degraded BC characteristics while introducing non-complementary MSI distribution among the legitimate receivers.

Let $\mathcal{X}$ represent the input set, while $\mathcal{Y}_1$, $\mathcal{Y}_2$, and $\mathcal{Z}$ denote the output sets corresponding to the strong receiver, weak receiver, and eavesdropper, respectively. The set $\mathcal{S}$ denotes the finite collection of channel states, which represent the CSI available at the transmitter. We use uppercase letters for discrete random variables and lowercase letters for their realizations. The entropy of a discrete random variable is denoted by $H(\cdot)$, and the mutual information between two random variables is represented by $I(\cdot; \cdot)$. Weak typicality is assumed in this work, and $\mathcal{T}_\epsilon^{(n)}(X)$ refers to the set of $\epsilon$-typical sequences $x^{(n)}$ over $n$ channel uses. The sequence of input random variables is denoted by $X_{\text{in}}$. A discrete memoryless degraded BC with two legitimate receivers, one eavesdropper, and channel states is formally defined by the sextuple
$(\mathcal{X}, \mathcal{S}, P(\mathcal{Y}_1, \mathcal{Y}_2, \mathcal{Z}|\mathcal{X}, \mathcal{S}), \mathcal{Y}_1, \mathcal{Y}_2, \mathcal{Z}),$
where the channel states $S_i \in \mathcal{S}$ are independent and identically distributed (i.i.d.) according to $p(s)$, and $P(\mathcal{Y}_1, \mathcal{Y}_2, \mathcal{Z}|\mathcal{X}, \mathcal{S})$ denotes the transition probabilities of the channel. In this work, we refer to $\mathcal{Y}_1$ as the strong receiver and $\mathcal{Y}_2$ as the weak receiver, consistent with their respective decoding capabilities. The transmitted messages are modeled as discrete random variables $M_1$ and $M_2$. 

The following definitions are introduced and will be utilized in subsequent sections of this paper.

\subsection{Definitions}

\textbf{Definition 1:} A $(n, 2^{nR_1}, 2^{nR_2}, \epsilon)$ code for a discrete memoryless two-receiver BC with non-causal CSI at the transmitter and complementary MSI at the receivers consists of two message sets $\mathcal{M}_1 := \{1, \dots, 2^{nR_1}\}$ and $\mathcal{M}_2 := \{1, \dots, 2^{nR_2}\}$, and the following three mappings:  
\begin{itemize}
    \item An encoder at the transmitter:
    \begin{equation}
        \text{enc.}: \mathcal{M}_1 \times \mathcal{M}_2 \times \mathcal{S} \to \mathcal{X}^n, \label{eq1}
    \end{equation}
    \item Decoders at the legitimate receivers:
    \begin{align}
        \text{dec.}(y_1): &\ \mathcal{Y}_1^n \times \mathcal{M}_2 \to \hat{\mathcal{M}}_1, \label{eq2} \\
        \text{dec.}(y_2): &\ \mathcal{Y}_2^n \times \mathcal{M}_1 \to \hat{\mathcal{M}}_2, \label{eq3}
    \end{align}
\end{itemize}
such that the average probability of error is defined as:
\begin{equation}
    P_{\text{Error}}^{(n)} := P\{(\hat{\mathcal{M}}_1, \hat{\mathcal{M}}_2) \neq (\mathcal{M}_1, \mathcal{M}_2)\} \leq \epsilon. \label{eq4}
\end{equation}

\textbf{Definition 2:} A rate pair $(R_1, R_2)$ is achievable if there exists an integer $n_0$ such that for all $n \geq n_0$, there exists a $(n, 2^{n(R_1-\mu)}, 2^{n(R_2-\mu)}, \epsilon)$ code for any $\mu > 0$.

\textbf{Definition 3:} The capacity region is defined as the closure of the union of all $\epsilon$-achievable rate pairs $(R_1, R_2)$.

The ignorance of the eavesdropper about the confidential messages $m_1 \in \mathcal{M}_1$ and $m_2 \in \mathcal{M}_2$ is measured using equivocation. The secrecy level of the confidential messages $\mathcal{M}_1$ and $\mathcal{M}_2$ is characterized in terms of equivocation rates, defined as:
\begin{align}
    R_{e1} &= \frac{1}{n} H(\mathcal{M}_1|Z^n), \label{eq5} \\
    R_{e2} &= \frac{1}{n} H(\mathcal{M}_2|Z^n), \label{eq6} \\
    R_{e12} &= \frac{1}{n} H(\mathcal{M}_1, \mathcal{M}_2|Z^n). \label{eq7}
\end{align}

\textbf{Definition 4:} A secrecy rate pair $(R_1, R_2) \in \mathbb{R}_+^2$ is achievable for the BC with complementary receiver SI if, for any $\delta > 0$, there exists $n(\delta) \in \mathbb{N}$ such that for all $n \geq n(\delta)$:
\begin{align}
    R_1 &\leq \delta + R_{e1}, \label{eq8} \\
    R_2 &\leq \delta + R_{e2}, \label{eq9} \\
    R_1 + R_2 &\leq \delta + R_{e12}, \label{eq10}
\end{align}
while ensuring that $P_{\text{Error}}^{(n)} \to 0$ as $n \to \infty$. These conditions guarantee perfect secrecy for each message individually and jointly.

\textbf{Definition 5:} The secrecy capacity region of the BC with non-causal CSI and complementary RMSI is the set of all achievable rate pairs $(R_1, R_2)$.

\textbf{Definition 6:} A $(n, 2^{nR_1}, 2^{nR_2}, \epsilon)$ code for a discrete memoryless two-receiver BC with non-causal CSI at the transmitter and non-complementary MSI at the receivers consists of two message sets $\mathcal{M}_1 := \{1, \dots, 2^{nR_1}\}$ and $\mathcal{M}_2 := \{1, \dots, 2^{nR_2}\}$, and the following mappings:
\begin{itemize}
    \item An encoder at the transmitter:
    \begin{equation}
        \text{enc.}: \mathcal{M}_1 \times \mathcal{M}_2 \times \mathcal{S} \to \mathcal{X}^n, \label{eq11}
    \end{equation}
    \item A decoder at each legitimate receiver:
    \begin{align}
        \text{dec.}(y_1): &\ \mathcal{Y}_1^n \to \hat{\mathcal{M}}_1, \label{eq12} \\
        \text{dec.}(y_2): &\ \mathcal{Y}_2^n \times \mathcal{M}_1 \to \hat{\mathcal{M}}_2, \label{eq13}
    \end{align}
\end{itemize}
such that the average probability of error is:
\begin{equation}
    P_{\text{Error}}^{(n)} := P\{(\hat{\mathcal{M}}_1, \hat{\mathcal{M}}_2) \neq (\mathcal{M}_1, \mathcal{M}_2)\} \leq \epsilon. \label{eq14}
\end{equation}

The definitions of achievable rates, capacity region, secrecy rate, and secrecy capacity region are identical to those stated in Definitions 2, 3, 4, and 5, respectively.

\section{Main Results}

In this section, we present the main results of our work, including an achievable secrecy rate region and an outer bound on the secrecy capacity region for the channels defined.

\textbf{Theorem 1.} An achievable secrecy rate region for a discrete memoryless two-receiver BC with non-causal CSI at the transmitter and complementary MSI at the receivers is given by the set of all rate pairs $(R_1, R_2) \in \mathbb{R}^2_+$ that satisfy:
\begin{equation}
    R_j \leq I(V; Y_j) - \max\{I(V; S), I(V; Z)\}, \quad j = 1, 2 \tag{15}
\end{equation}
where the random variables follow the joint probability distribution: $p(s, v, x, y_1, y_2, z) = p(s)p(v|s)p(x|v, s)p(y_1, y_2, z|x, s).$

\textbf{Proof.} The achievable secrecy rate region is derived using a combination of random coding, superposition coding, Marton coding, and an extension of Gel'fand-Pinsker coding. The intuition behind (15) arises from considering the random distribution of $2^n[I(V; Y_j)]$ sequences into $2^{nR_j}$ bins, ensuring that the bit rate of each bin sequence exceeds $\max\{I(V; S), I(V; Z)\}$. The detailed proof is outlined below.

\textbf{Random Codebook $\mathcal{C}^n$:}
We first generate $2^{n[I(V; Y_j) - \epsilon_{VY_j}]}$ independent and identically distributed (i.i.d.) sequences $v^n$, where the distribution of $v^n$ is $P_{v^n}(v^n) = \prod_{i=1}^n P_V(v_i)$. These sequences are then randomly distributed into $2^{nR_j}$ bins, with each bin containing $2^{n[\max\{I(V; S), I(V; Z)\} + \epsilon_{VSZ}]}$ sequences. The rate $R_j$ is thus:
\begin{equation}
R_j = I(V; Y_j) - \max\{I(V; S), I(V; Z)\} - \epsilon_{VY_j} - \epsilon_{VSZ}.
\end{equation}
Each bin is indexed by $k_j \in \{1, 2, \ldots, 2^{nR_j}\}$, and the sequences within each bin are further divided into $2^{n[I(V; Z) - \epsilon_{VZ}]}$ sub-bins. Let $W$ denote the sub-bin index of a sequence $v^n$. Then, $W$ takes values from:
\begin{equation}
    W \in \{1, 2, \ldots, 2^{n[\max\{I(V; S), I(V; Z)\} - I(V; Z) + \epsilon_{VSZ} + \epsilon_{VZ}]}\}.
\end{equation}

Finally, fix an input distribution $Q_X(x)$ and construct $x^n(m_1, m_2, m_s)$ for $m_j \in M_j = \{1, \ldots, 2^{nR_j}\}$, $j = 1, 2$, and $m_s \in M_s = \{1, \ldots, 2^{nR_s}\}$.

\textbf{Encoding}:
To transmit message $k_j$ given the state information $s^n$, the transmitter searches for a sequence $v^n(k_j)$ in bin $k_j$ such that $(v^n(k_j), s^n)$ is jointly typical:
\begin{equation}
(v^n(k_j), s^n) \in \mathcal{T}_{V(n,S)}(\epsilon).
\end{equation}
If no such sequence exists, the transmitter randomly selects a sequence from bin $k_j$ and transmits the corresponding $x^n(k_j)$, which is generated according to:
\begin{equation}
    P_{X^n|V^n,S^n}(x^n(k_j)|v^n(k_j), s^n) = \prod_{i=1}^n P_{X|V,S}(x_i|v_i, s_i).
\end{equation}

\textbf{Decoding}:
The legitimate receivers decode the received signals $y_1^n$ and $y_2^n$ using their respective decoding strategies.

\textbf{First Decoder (Dec. $y_1$):} Given $y_1^n$ and its message index $m_2$, the receiver determines the pair $(\hat{m}_1, \hat{m}_s)$ if it is unique and satisfies:
$(x^n(\hat{m}_1, m_2, \hat{m}_s), y_1^n) \in \mathcal{T}_{X,Y_1}(\epsilon).
$ Otherwise, an error is declared.

\textbf{Second Decoder (Dec. $y_2$):} Given $y_2^n$ and its message index $m_1$, the receiver determines the pair $(\hat{m}_2, \hat{m}_s)$ if it is unique and satisfies:
$(x^n(m_1, \hat{m}_2, \hat{m}_s), y_2^n) \in \mathcal{T}_{X,Y_2}(\epsilon).
$ Otherwise, an error is declared.

The encoding and decoding strategies used here are extensions of those presented in [8, 20, 30]. Consequently, the secrecy rates $R_j$ for $j = 1, 2$ from the transmitter to the legitimate receivers are shown to be achievable.

\textbf{Analysis of Probability of Error:} For the legitimate receivers $Y_1$ and $Y_2$, we have the following errors:
\begin{itemize}
    \item $\epsilon^S(k_j)$: In the encoding process, given $s^n$ and message $k_j$, there is no sequence $v^n$ in the bin $k_j$ that is jointly typical with $s^n$.
    \item $\epsilon^{y_1}(k_1)$: In the encoding process, there is no sequence $v^n$ that is jointly typical with the received sequence $y_1^n$.
    \item $\epsilon^{y_1}(k_1')$: In the encoding process, there is a sequence $v^n(k_1')$ in bin $k_1', k_1' \neq k_1$, that is jointly typical with the received sequence $y_1^n$.
    \item $\epsilon^{y_2}(k_2)$: In the encoding process, there is no sequence $v^n$ that is jointly typical with the received sequence $y_2^n$.
    \item $\epsilon^{y_2}(k_2')$: In the encoding process, there is a sequence $v^n(k_2')$ in bin $k_2', k_2' \neq k_2$, that is jointly typical with the received sequence $y_2^n$.
\end{itemize}

We first analyze the probability of $\epsilon^S(k_j)$. Sequences $v^n$ and $s^n$ are independent by the code generating process. For $n$ sufficiently large, the probability that a pair $(v^n, s^n)$ is jointly typical is larger than $(1 - \epsilon)2^{n[I(V;S)+3\epsilon]}$. Thus, we have the following inequality:
\begin{equation}
    \Pr\{(v^n, s^n) \in \mathcal{I}_V^{(n,S)}(\epsilon)\} \geq (1 - \epsilon)^{2n[I(V;S) + 3\epsilon]},
\end{equation}
\begin{equation}
    \Pr\{(v^n, s^n) \notin \mathcal{I}_V^{(n,S)}(\epsilon)\} \leq 1 - (1 - \epsilon)^{2n[I(V;S) + 3\epsilon]},
\end{equation}
\begin{multline}
    \Pr\{\epsilon^S(k_j) \mid (\hat{\mathcal{M}}_1, \hat{\mathcal{M}}_2) = (\mathcal{M}_1, \mathcal{M}_2)\}
\leq  \big[1 - (1 - \epsilon)^{2n[I(V;S) + 3\epsilon]}\big] \cdot 2^{n[\max\{I(V;S), I(V;Z)\} + \epsilon_{VSZ}]}
    \leq\\ \exp\big\{-(1 - \epsilon)^{2n[I(V;S) + 3\epsilon]} \cdot 2^{n[\max\{I(V;S), I(V;Z)\} + \epsilon_{VSZ}]}\big\}
    \leq \exp\big\{-(1 - \epsilon)^{2n[\max\{I(V;S), I(V;Z)\} - I(V;S) + \epsilon_{VSZ} - 3\epsilon]}\big\}
    \leq \delta^3,
\end{multline}
where the first inequality in (21) follows from the fact that there are 
\( 2^{n[\max\{I(V;S), I(V;Z)\} + \epsilon_{VSZ}]} \) codewords in a bin. The second inequality in (21) follows from
the \( e^a \geq 1 + a \). As shown above, for a given \( \epsilon \) and
arbitrarily small \( \delta \), there exists \( \epsilon_{VSZ} \) such that when \( \epsilon_{VSZ} > 3\epsilon \),
both \( \Pr\{(v^n, s^n) \in \mathcal{I}_V^{(n,S)}(\epsilon)\} \geq (1 - \epsilon)^{2n[I(V;S) + 3\epsilon]} \) and 
\( \Pr\{\epsilon^S(k_j) \mid (\hat{\mathcal{M}}_1, \hat{\mathcal{M}}_2) = (\mathcal{M}_1, \mathcal{M}_2)\} \leq \delta^3 \). 

Now, we analyze the probability of \( \epsilon^{y_1}(k_1) \) and \( \epsilon^{y_2}(k_2) \). If the event \( \epsilon^S(k_j) \) does
not occur, which means that there is a sequence \( v^n(k_j) \) in bin \( k_j \)
and a sequence \( x^n(k_j) \) such that \( (v^n(k_j), x^n(k_j), s^n) \) is jointly
typical, then each of \( (v^n(k_1), x^n(k_1), s^n, y_1^n) \) and 
\( (v^n(k_2), x^n(k_2), s^n, y_2^n) \) will be jointly typical with high
probability. For a given \( \epsilon \) and arbitrarily small \( \delta \), we have
\begin{equation}
\Pr\{(v^n, y_1^n) \in \mathcal{I}_V^{(n,Y_1)}(\epsilon)\} + \Pr\{(v^n, y_2^n) \in \mathcal{I}_V^{(n,Y_2)}(\epsilon)\} \geq 1 - \delta^3,
\end{equation}
which implies that
\begin{equation}
\Pr\{\epsilon^{y_1}(k_1), \mathcal{M}_2 \mid \epsilon^S(k_1)^C, \hat{\mathcal{M}}_1 = \mathcal{M}_1\}
+ \Pr\{\epsilon^{y_2}(k_2), \mathcal{M}_1 \mid \epsilon^S(k_2)^C, \hat{\mathcal{M}}_2 = \mathcal{M}_2\} \leq \delta^3.
\end{equation}

Now, we analyze the probability of \( \epsilon^{y_1}(k_1') \) and \( \epsilon^{y_2}(k_2') \). If we say that \( \epsilon^{{y_1}^*}(k_1') \) and \( \epsilon^{{y_2}^*}(k_2') \) occur when some other \( v_N \) is
jointly typical with \( y_1^n \) and \( y_2^n \), respectively, then it is clear that
\begin{equation}
\Pr\{\epsilon^{y_1}(k_1'), \mathcal{M}_2 \mid \epsilon^S(k_1)^C, \hat{\mathcal{M}}_1 = \mathcal{M}_1\}
\leq \Pr\{\epsilon^{{y_1}^*}(k_1'), \mathcal{M}_2 \mid \epsilon^S(k_1)^c, \hat{\mathcal{M}}_1 = \mathcal{M}_1\} 
\end{equation}
and
\begin{equation}
\Pr\{\epsilon^{y_2}(k_2'), \mathcal{M}_1 \mid \epsilon^S(k_2)^C, \hat{\mathcal{M}}_2 = \mathcal{M}_2\}
\leq \Pr\{\epsilon^{{y_2}^*}(k_2'), \mathcal{M}_1 \mid \epsilon^S(k_2)^C, \hat{\mathcal{M}}_2 = \mathcal{M}_2\}. 
\end{equation}

But a sequence \( v^n \), different from \( v^n(k_j) \), being jointly typical
with \( y_1^n \) and \( y_2^n \) has a probability of at most \( 2^{-n[I(V;Y_j) - 3\epsilon]} \). Since
there are only \( 2^{n[I(V;Y_j) - \epsilon_{VY_j}]} \) other sequences, for a given \( \epsilon \) and
arbitrarily small \( \delta \), there exists \( \epsilon_{VS} \) such that
\begin{multline}
\Pr\{\epsilon^{y_1}(k_1), \mathcal{M}_2 \mid \epsilon^S(k_1)^C, \hat{\mathcal{M}}_1 = \mathcal{M}_1\}
+ \Pr\{\epsilon^{y_2}(k_2), \mathcal{M}_1 \mid \epsilon^S(k_2)^C, \hat{\mathcal{M}}_2 = \mathcal{M}_2\}
\\
\leq \Pr\{\epsilon^{{y_1}^*}(k_1'), \mathcal{M}_2 \mid \epsilon^S(k_1)^C, \hat{\mathcal{M}}_1 = \mathcal{M}_1\}
+ \Pr\{\epsilon^{{y_2}^*}(k_2'), \mathcal{M}_1 \mid \epsilon^S(k_2)^C, \hat{\mathcal{M}}_2 = \mathcal{M}_2\}
\\
\leq \sum_{v_n \neq v_n(k_j)} 2^{-n[I(V;Y_j) - 3\epsilon]} \leq 
(2^{n[I(V;Y_j) - \epsilon_{VY_j}]} - 1) .2^{-n[I(V;Y_j) - 3\epsilon]}  \leq 2^{-n[\epsilon_{VY_j} - 3\epsilon]} \leq \delta^3. 
\end{multline}

All error events are arbitrarily small, as shown above. By the
union bound on these probabilities of error, for \( \epsilon_{VY_j}, \epsilon_{VSZ} > 3\epsilon \),
the average probability of error is as follows:
\begin{multline}
P_{\text{Error}}(n)
=
\frac{1}{2^{n(R_1 + R_2)}} \sum_{j=1}^{2^{nR_1}} \sum_{j=1}^{2^{nR_2}} \Pr\{(\hat{\mathcal{M}}_1, \hat{\mathcal{M}}_2) \neq (\mathcal{M}_1, \mathcal{M}_2) \mid (\mathcal{M}_1, \mathcal{M}_2) = (\mathcal{M}_1, \mathcal{M}_2)\}\leq \\ \frac{1}{2^{n(R_1 + R_2)}} 
\sum_{j=1}^{2^{nR_1}} \sum_{j=1}^{2^{nR_2}} 
\Big[ \Pr\{\epsilon^S(k_j) \mid (\hat{\mathcal{M}}_1, \hat{\mathcal{M}}_2) = (\mathcal{M}_1, \mathcal{M}_2)\} \\+ \Pr\{\epsilon^{y_1}(k_1), \mathcal{M}_2 \mid \epsilon^S(k_1)^C, \hat{\mathcal{M}}_1 = \mathcal{M}_1\} + \Pr\{\epsilon^{y_2}(k_2), \mathcal{M}_1 \mid \epsilon^S(k_2)^C, \hat{\mathcal{M}}_2 = \mathcal{M}_2\}
\Big]\\+ \Pr\{\epsilon^{{y_1^*}(k_1')}, \mathcal{M}_2 \mid \epsilon^S(k_1)^C, \hat{\mathcal{M}}_1 = \mathcal{M}_1\} \notag \
+ \Pr\{\epsilon^{{y_2^*}(k_2')}, \mathcal{M}_1 \mid \epsilon^S(k_2)^C, \hat{\mathcal{M}}_2 = \mathcal{M}_2\}  
\end{multline}
\begin{equation}
\leq \frac{1}{2^n(R_1+R_2)} \sum_{j=1}^{2^{nR_1}} \sum_{j=1}^{2^{nR_2}} \left[\delta_3 + \delta_3 + \delta_3\right] = \delta. 
\end{equation}
In the above inequalities, $\delta \to 0$ as $n \to \infty$, so $P_{\text{Error}}(n) \to 0$.  

\textbf{Equivocation Computation:} Now, we demonstrate that the equivocation rates \( R_{e1} \), \( R_{e2} \), and \( R_{e12} \), achieved by the proposed codebook construction, satisfy the secrecy conditions specified in (8), (9), and (10), respectively. Specifically, we focus on proving the secrecy condition for the individual message \( \mathcal{M}_1 \) (condition (8) in this study). The proofs for conditions (9) and (10) follow similarly and are omitted here for brevity. To establish the requirement in (8), we proceed as follows:
\begin{align}
nR_{e1} &= H(\mathcal{M}_1 \mid Z^n) = H(\mathcal{M}_1, Z^n) - H(Z^n) \\
&= H(\mathcal{M}_1, W, Z^n) - H(W \mid \mathcal{M}_1, Z^n) - H(Z^n) \\
&= H(\mathcal{M}_1, W, V^n, Z^n) - H(V^n \mid \mathcal{M}_1, W, Z^n) \\
&- H(W \mid \mathcal{M}_1, Z^n) \notag \quad - H(Z^n) = H(V^n \mid Z^n)  \\
&- H(V^n \mid \mathcal{M}_1, W, Z^n)- H(W \mid \mathcal{M}_1, Z^n)\\
&\geq H(V^n \mid Z^n) - H(V^n \mid \mathcal{M}_1, W, Z^n) - \log|W| \\
&- H(V^n \mid Y^n) = n[I(V; Y) - I(V; Z)]  \\
&- H(V^n \mid \mathcal{M}_1, W, Z^n) \notag \quad - n[\max\{I(V; S), I(V; Z)\} \\
& - I(V; Z) 
+ \epsilon_{VSZ} + \epsilon_{VZ}] = nR_1 - \\
&n(\epsilon_{VSZ} + \epsilon_{VZ} + \frac{1}{n} H(V^n \mid \mathcal{M}_1, W, Z^n)) \\
&= n(R_1 - \delta). 
\end{align}
Here, (32) follows from the fact that $H(W \mid \mathcal{M}_1, Z^n) \leq H(W) \leq \log|W|$, and (33) follows from the fact that $\log|W| = n[\max\{I(V; S), I(V; Z)\} - I(V; Z) + \epsilon_{VSZ} + \epsilon_{VZ}]$. Thus, it has been shown that the desired secrecy condition (8) is satisfied. This concludes the proof that any rate pairs $(R_1, R_2)$ are achievable. \qed

\textbf{Theorem 2.} An outer bound on the secrecy capacity region for a discrete memoryless 2-receiver BC with non-causal CSI at the transmitter and complementary MSI at receivers is given by the set of all rate pairs $(R_1, R_2) \in \mathbb{R}_+^2$ that satisfy the following conditions:
\begin{align}
R_1 &\leq I(V; Y_1) - \max\{I(V; S), I(V; Z)\},  \\
R_2 &\leq I(V; Y_2) - \max\{I(V; S), I(V; Z)\}, \\
R_1 + R_2 &\leq \min\{I(V; Y_1 \mid U) + I(V; Y_2 \mid U) - I(V; Z \mid U), \notag \\
&\quad I(V; Y_1) + I(V; Y_2) - I(V; Z)\}. 
\end{align}
Here, \( U - V - X - (Y_1, Y_2, Z) \) represents the relevant random variables.

\textbf{Remark 1:} If the secrecy constraints are removed by setting \( Z = 0 \), the above rate region reduces to the inverse part of the capacity region for the 2-user degraded BC with state information at the transmitter and MSI at the receivers.

\textbf{Remark 2:} By removing the state information constraints, i.e., setting \( S = 0 \), the resulting rate region reduces to an outer bound on the secrecy capacity region of a broadcast channel with two legitimate receivers, one external eavesdropper, and MSI available at the receivers.

\textbf{Proof:} To derive the desired outer bound on the secrecy capacity region, we employ a version of Fano’s lemma applicable to the BC with receiver SI, expressed as:  
\begin{align}
H(\mathcal{M}_1 \mid Y_1^n, \mathcal{M}_2) &\leq n\epsilon_1(n),  \\
H(\mathcal{M}_2 \mid Y_2^n, \mathcal{M}_1) &\leq n\epsilon_2(n), 
\end{align}
where $\epsilon_1(n)$ and $\epsilon_2(n)$ satisfy $\epsilon_1(n), \epsilon_2(n) \to 0$ as $n \to \infty$.

To facilitate the proof, we define the following auxiliary random variables:  
\begin{align}
U_i &\triangleq (Y_1^{i-1}, Y_2^{i-1}, Z_{i+1}^n, S_{i+1}^n),  \\
V_i &\triangleq (\mathcal{M}_1, \mathcal{M}_2, U_i), 
\end{align}
which satisfy the Markov chain condition: $U_i - V_i - X_i - (Y_1, Y_2, Z).$ 

Let $\mathcal{M}_1$ and $\mathcal{M}_2$ be independent random variables representing the transmitted messages. The rate $R_1$ can be bounded as follows:
\begin{align}
&nR_1 \leq H(M_1 | Z^n) + n\delta  \\
&\leq H(M_1) + n\delta = H(M_1 | M_2) + n\delta  \\
&= H(M_1 | Y_1^n, M_2) + I(M_1; Y_1^n | M_2) + n\delta  \\
&\leq \sum_{i=1}^n I(M_1; Y_{1,i} | M_2, Y_1^{i-1}) + n\epsilon_1(n) + n\delta \\
&\nonumber\leq \sum_{i=1}^n I(M_1, M_2, Y_1^{i-1}, Y_2^{i-1}, Z_{i+1}^n; Y_{1,i}) \\&+ n(\epsilon_1(n) + \delta) \ \\
&\nonumber= \sum_{i=1}^n I(M_1, M_2, Y_1^{i-1}, Y_2^{i-1}, Z_{i+1}^n, S_{i+1}; Y_{1,i})  -\\
& \nonumber\sum_{i=1}^n I(S_{i+1}; Y_{1,i} | M_1, M_2, Y_1^{i-1}, Y_2^{i-1}, Z_{i+1}^n)\\&+ n(\epsilon_1(n) + \delta) \\
&\nonumber= \sum_{i=1}^n I(M_1, M_2, Y_1^{i-1}, Y_2^{i-1}, Z_{i+1}^n, S_{i+1}; Y_{1,i})  -\\\nonumber
& \sum_{i=1}^n I(Y_1^{i-1}; S_i | M_1, M_2, S_{i+1}, Y_2^{i-1}, Z_{i+1}^n)\\&+ n(\epsilon_1(n) + \delta) \\
&\nonumber= \sum_{i=1}^n I(M_1, M_2, Y_1^{i-1}, Y_2^{i-1}, Z_{i+1}^n, S_{i+1}; Y_{1,i}) -\\
& \nonumber\sum_{i=1}^n I(M_1, M_2, S_{i+1}, Y_1^{i-1}, Y_2^{i-1}, Z_{i+1}^n; S_i)\\&+ n(\epsilon_1(n) + \delta) \\
&= \sum_{i=1}^n I(V_i; Y_{1,i}) - \sum_{i=1}^n I(V_i; S_i)+ n(\epsilon_1(n) + \delta), 
\end{align}
where the first inequality follows from the perfect secrecy condition (8) and (47) is due to Fano’s inequality. (49) follows from the Csiszár–Körner identity. Finally, the last equality comes from the definition of the auxiliary random variables (42). Accordingly, from the perfect secrecy condition (9), $R_2$ can be bounded as follows:
\begin{equation}
nR_2 \leq \sum_{i=1}^n I(V_i; Y_{2,i}) - \sum_{i=1}^n I(V_i; S_i) + n(\epsilon_2(n) + \delta). 
\end{equation}
Also, we have:
\begin{align}
&nR_1 \leq H(M_1 | Z^n) + n\delta  \\
&\leq H(M_1) + n\delta = H(M_1 | M_2) + n\delta. 
\\
&= H(M_1|Y_1^n, M_2) + I(M_1; Y_1^n|M_2) + n\delta  \\
&\leq \sum_{i=1}^n I(M_1; Y_{1,i}|M_2, Y_1^{i-1}) + n\epsilon_1(n) + n\delta \\
&\leq \sum_{i=1}^n I(M_1, M_2, Y_1^{i-1}, Y_2^{i-1}, S_i^{n+1}; Y_{1,i}) + n(\epsilon_1(n) + \delta)  \\
&= \sum_{i=1}^n I(M_1, M_2, Y_1^{i-1}, Y_2^{i-1}, Z_i^{n+1}, S_i^{n+1}; Y_{1,i}) \notag - \\
&\sum_{i=1}^n I(Z_i^{n+1}; Y_{1,i}|M_1, M_2, Y_1^{i-1}, Y_2^{i-1}, S_i^{n+1}) + n(\epsilon_1(n) + \delta)  \\
&= \sum_{i=1}^n I(M_1, M_2, Y_1^{i-1}, Y_2^{i-1}, Z_i^{n+1}, S_i^{n+1}; Y_{1,i}) \notag -\\
&\sum_{i=1}^n I(Y_1^{i-1}; Z_i|M_1, M_2, S_i^{n+1}, Y_2^{i-1}, Z_i^{n+1}) + n(\epsilon_1(n) + \delta)  \\
&= \sum_{i=1}^n I(M_1, M_2, Y_1^{i-1}, Y_2^{i-1}, Z_i^{n+1}, S_i^{n+1}; Y_{1,i}) \notag -\\
& \sum_{i=1}^n I(M_1, M_2, Y_1^{i-1}, Y_2^{i-1}, Z_i^{n+1}, S_i^{n+1}; Z_i) + n(\epsilon_1(n) + \delta)  \\
&= \sum_{i=1}^n I(V_i; Y_{1,i}) - \sum_{i=1}^n I(V_i; Z_i) + n(\epsilon_1(n) + \delta). 
\end{align}
The first inequality arises from the perfect secrecy constraint stated in (8), while the transition leading to (57) is justified by Fano’s inequality. Equation (16) is derived using the Csiszár–Körner identity, and the final equality follows directly from the definition of the auxiliary random variables introduced in (42). Accordingly, based on the perfect secrecy requirement outlined in (9), the rate $R_2$ admits the following upper bound:
\begin{align}
nR_2 &\leq \sum_{i=1}^n I(V_i; Y_{2,i}) - \sum_{i=1}^n I(V_i; Z_i) + n(\epsilon_2(n) + \delta). 
\end{align}
To bound the sum rate $R_1 + R_2$, we use the fact that $M_1$ and $M_2$ are independent messages. So, $R_1 + R_2$ can be bounded as follows:
\begin{align}
&n(R_1 + R_2) \leq H(M_1, M_2|Z^n) + n\delta
\\
&\nonumber= H(M_1, M_2|Z^n) - H(M_1|Y_1^n, M_2) + H(M_1|Y_1^n, M_2) \\& - H(M_2|Y_2^n, M_1) + H(M_2|Y_2^n, M_1) + n\delta  \\
&\leq H(M_1, M_2|Z^n) - H(M_1|Y_1^n, M_2) \notag \\
&\quad - H(M_2|Y_2^n, M_1) + n(\epsilon_1(n) + \epsilon_2(n) + \delta)  \\
&\nonumber= H(M_1|M_2) + H(M_2|M_1) - H(M_1, M_2) \\
&+ H(M_1, M_2|Z^n) \notag \quad - H(M_1|Y_1^n, M_2) \\
&- H(M_2|Y_2^n, M_1) + n(\epsilon_1(n) + \epsilon_2(n) + \delta). \\& \nonumber
= I(M_1; Y_1^n | M_2) + I(M_2; Y_2^n | M_1)
- I(M_1, M_2; Z^n) + \\& n(\epsilon_1(n) + \epsilon_2(n) + \delta) \\&\nonumber
\leq I(M_1, M_2; Y_1^n) + I(M_1, M_2; Y_2^n)
- I(M_1, M_2; Z^n) + \\& n(\epsilon_1(n) + \epsilon_2(n) + \delta) \\&\nonumber
= \sum_{i=1}^n I(M_1, M_2; Y_{1,i} | Y_1^{i-1})
+ \sum_{i=1}^n I(M_1, M_2; Y_2^{i} | Y_2^{i-1})\nonumber\\&
- \sum_{i=1}^n I(M_1, M_2; Z_i | Z_i^n)
+ n(\epsilon_1(n) + \epsilon_2(n) + \delta) \\&\nonumber
= \sum_{i=1}^n I(M_1, M_2, Y_2^{i-1}, Z_i^n, S_i^n; Y_1^{i} | Y_1^{i-1})
\\&\nonumber- \sum_{i=1}^n I(Y_2^{i-1}, Z_i^n, S_i^n; Y_{1,i} | M_1, M_2, Y_1^{i-1})
+ \\&\nonumber\sum_{i=1}^n I(M_1, M_2, Y_1^{i-1}, Z_i^n, S_i^n; Y_2^{i} | Y_2^{i-1}) \\&\nonumber
- \sum_{i=1}^n I(Y_1^{i-1}, Z_i^n, S_i^n; Y_2^{i} | M_1, M_2, Y_2^{i-1})
\\&\nonumber- \sum_{i=1}^n I(M_1, M_2, Y_1^{i-1}, Y_2^{i-1}, S_i^n; Z_i | Z_i^n)
\\&+ \nonumber\sum_{i=1}^n I(Y_1^{i-1}, Y_2^{i-1}, S_i^n; Z_i | M_1, M_2, Z_i^n)
\\&+ n (\epsilon_1(n) + \epsilon_2(n) + \delta) \\&\nonumber
\leq \sum_{i=1}^n I(M_1, M_2, Y_2^{i-1}, Z_i^n, S_i^n; Y_1^{i} | Y_1^{i-1})
\\&\nonumber- \sum_{i=1}^n I(Y_2^{i-1}, Z_i^n, S_i^n; Y_1^{i} | M_1, M_2, Y_1^{i-1})
\\&+ \nonumber\sum_{i=1}^n I(M_1, M_2, Y_1^{i-1}, Z_i^n, S_i^n; Y_2^{i} | Y_2^{i-1}) \\&\nonumber
- \sum_{i=1}^n I(Y_1^{i-1}, Z_i^n, S_i^n; Y_2^{i} | M_1, M_2, Y_2^{i-1})
\\&\nonumber- \sum_{i=1}^n I(M_1, M_2, Y_1^{i-1}, Y_2^{i-1}, S_i^n; Z_i | Z_i^n)
+\\& \nonumber\sum_{i=1}^n I(Y_1^{i-1}, Y_2^{i-1}; S_i, Z_i | M_1, M_2, S_i^n, Z_i^n)
\\&+ n(\epsilon_1(n) + \epsilon_2(n) + \delta) 
\\
&\nonumber
= \sum_{i=1}^n I(M_1, M_2; Y_1^{i} | Y_1^{i-1}, Y_2^{i-1}, Z_i^n, S_i^n)
\\&\nonumber+ \sum_{i=1}^n I(Y_2^{i-1}, Z_i^n, S_i^n; Y_1^{i} | Y_1^{i-1})\\&\nonumber
    + \sum_{i=1}^{n} I(M_1, M_2; Y_2^{i} \mid Y_1^{i-1}, Y_2^{i-1}, Z_{i}^{n+1}, S_{i}^{n+1}) \\
    &\nonumber+ \sum_{i=1}^{n} I(Y_1^{i-1}, Z_{i}^{n+1}, S_{i}^{n+1}; Y_2^{i} \mid Y_2^{i-1}) \nonumber \\
    &- \sum_{i=1}^{n} I(M_1, M_2; Z_i \mid Y_1^{i-1}, Y_2^{i-1}, S_{i}^{n+1}, Z_{i}^{n+1}) \nonumber \\
    &- \sum_{i=1}^{n} I(Y_1^{i-1}, Y_2^{i-1}, S_{i}^{n+1}; Z_i \mid Z_{i}^{n+1}) \nonumber \\
    &+ n(\epsilon_1(n) + \epsilon_2(n) + \delta)  \\
    &\leq \sum_{i=1}^{n} I(M_1, M_2; Y_1^{i} \mid Y_1^{i-1}, Y_2^{i-1}, Z_{i}^{n+1}, S_{i}^{n+1}) \nonumber \\
    &+ \sum_{i=1}^{n} I(Y_2^{i-1}, Z_{i}^{n+1}, S_{i}^{n+1}; Y_1^{i} \mid Y_1^{i-1}) \nonumber \\
    &+ \sum_{i=1}^{n} I(M_1, M_2; Y_2^{i} \mid Y_1^{i-1}, Y_2^{i-1}, Z_{i}^{n+1}, S_{i}^{n+1}) \nonumber \\
    &+ \sum_{i=1}^{n} I(Y_1^{i-1}, Z_{i}^{n+1}, S_{i}^{n+1}; Y_2^{i} \mid Y_2^{i-1}) \nonumber \\
    &- \sum_{i=1}^{n} I(M_1, M_2; Z_i \mid Y_1^{i-1}, Y_2^{i-1}, S_{i}^{n+1}, Z_{i}^{n+1}) \nonumber \\
    &- \sum_{i=1}^{n} I(Y_1^{i-1}, Y_2^{i-1}; S_i, Z_i \mid S_{i}^{n+1}, Z_{i}^{n+1}) \nonumber \\
    &+ n(\epsilon_1(n) + \epsilon_2(n) + \delta)  \\
    &= \sum_{i=1}^{n} I(M_1, M_2; Y_1^{i} \mid Y_1^{i-1}, Y_2^{i-1}, Z_{i}^{n+1}, S_{i}^{n+1}) \nonumber \\
    &+ \sum_{i=1}^{n} I(M_1, M_2; Y_2^{i} \mid Y_1^{i-1}, Y_2^{i-1}, Z_{i}^{n+1}, S_{i}^{n+1}) \nonumber \\
    &- \sum_{i=1}^{n} I(M_1, M_2; Z_i \mid Y_1^{i-1}, Y_2^{i-1}, S_{i}^{n+1}, Z_{i}^{n+1}) \nonumber \\
    &+ n(\epsilon_1(n) + \epsilon_2(n) + \delta)  \\
    & = \sum_{i=1}^{n} I(V_i; Y_1^{i} \mid U_i) + \sum_{i=1}^{n} I(V_i; Y_2^{i} \mid U_i) \nonumber \\
    &- \sum_{i=1}^{n} I(V_i; Z_i \mid U_i) + n(\epsilon_1(n) + \epsilon_2(n) + \delta).
\end{align}
The first inequality is a consequence of the perfect secrecy condition given in (10), while the derivation of (72) follows from Fano’s inequality. Lemma 1 presented below as a tailored version of the Csiszár–Körner sum identity for the considered setting—is utilized in steps (79) and (81). Lastly, the final equality directly follows from the definitions of the auxiliary random variables in (48) and (49).

\textbf{Lemma 1.} We have the following identities:
\begin{align}
    &\nonumber\sum_{i=1}^{n} I(Y_2^{i-1}, Z_{i}^{n+1}, S_{i}^{n+1}; Y_{1,i} \mid Y_1^{i-1}) \\&\nonumber+
    \sum_{i=1}^{n} I(Y_1^{i-1}, Z_{i}^{n+1}, S_{i}^{n+1}; Y_2^{i} \mid Y_2^{i-1})\\&=
\sum_{i=1}^{n} I(Y_1^{i-1}, Y_2^{i-1}; S_i, Z_i | Z_{i+1}^{n}, S_{i+1}^{n}) 
\end{align}
and
\begin{align}
& \nonumber\sum_{i=1}^{n} I(Y_2^{i-1}, Z_{i+1}^{n}, S_{i+1}^{n}; Y_1^{i} | M_1, M_2, Y_1^{i-1})\\& \nonumber
+ \sum_{i=1}^{n} I(Y_1^{i-1}, Z_{i+1}^{n}, S_{i+1}^{n}; Y_2^{i} | M_1, M_2, Y_2^{i-1})\\& 
= \sum_{i=1}^{n} I(Y_1^{i-1}, Y_2^{i-1}; S_i, Z_i | M_1, M_2, Z_{i+1}^{n}, S_{i+1}^{n}). 
\end{align}

{\bf{Proof.}} To prove the first identity, according to Lemma 7 in [8], we use the chain rule to represent the mutual information terms on the left-hand side of (77) as follows:
\begin{align}
&\nonumber I(Y_2^{i-1}, Z_{i+1}^{n}, S_{i+1}^{n}; Y_{1,i} | Y_1^{i-1}) \\&\nonumber
+ I(Y_1^{i-1}, Z_{i+1}^{n}, S_{i+1}^{n}; Y_2^{i} | Y_2^{i-1}) \\&\nonumber
= \sum_{j=i+1}^{n} \big[ I(Z_j, S_j; Y_1^{i} | Y_1^{i-1}, Y_2^{i-1}, Z_{j+1}^{n}, S_{j+1}^{n}) 
\\&+ I(Z_j, S_j; Y_2^{i} | Y_1^{i-1}, Y_2^{i-1}, Z_{j+1}^{n}, S_{j+1}^{n}) \big] \\&\nonumber
= \sum_{j=i+1}^{n} \big[ I(Z_j; Y_1^{i} | Y_1^{i-1}, Y_2^{i-1}, Z_{j+1}^{n}, S_{j+1}^{n}) 
\\&\nonumber+ I(Z_j; Y_{2,i} | Y_1^{i-1}, Y_2^{i-1}, Z_{j+1}^{n}, S_{j+1}^{n}) 
+ \\&\nonumber I(S_j; Y_{1,i} | Y_1^{i-1}, Y_2^{i-1}, Z_{j+1}^{n}, S_{j+1}^{n}, Z_j) 
\\&+ I(S_j; Y_2^{i} | Y_1^{i-1}, Y_2^{i-1}, Z_{j+1}^{n}, S_{j+1}^{n}, Z_j) \big] 
\\&\nonumber
= \sum_{j=i+1}^{n} \big[ I(Z_j; Y_1^{i} | Y_1^{i-1}, Y_2^{i-1}, Z_{j+1}^{n}, S_{j+1}^{n}) 
\\&\nonumber+ I(Z_j; Y_2^{i} | Y_1^{i-1}, Y_2^{i-1}, Z_{j+1}^{n}, S_{j+1}^{n}) 
+\\&\nonumber I(S_j; Y_1^{i} | Y_1^{i-1}, Y_2^{i-1}, Z_{j+1}^{n}, S_{j+1}^{n}) 
+ \\&\nonumber I(S_j; Y_2^{i} | Y_1^{i-1}, Y_2^{i-1}, Z_{j+1}^{n}, S_{j+1}^{n}) \big].
\end{align}
and on the right-hand side as:
\begin{align}
&\nonumber
I(Y_1^{i-1}, Y_2^{i-1}; S_i, Z_i | S_{i+1}^{n}, Z_{i+1}^{n}) 
\\&= \sum_{j=1}^{i-1} I(Y_1^{j}, Y_2^{j}; S_i, Z_i | Y_1^{j-1}, Y_2^{j-1}, S_{i+1}^{n}, Z_{i+1}^{n}) \\&
= \sum_{j=1}^{i-1} \big[ I(Y_1^{j}; S_i, Z_i | Y_1^{j-1}, Y_2^{j-1}, S_{i+1}^{n}, Z_{i+1}^{n}) 
\\&\nonumber+ I(Y_2^{j}; S_i, Z_i | Y_1^{j}, Y_1^{j-1}, Y_2^{j-1}, S_{i+1}^{n}, Z_{i+1}^{n}) \big]. 
\\&
= \sum_{j=1}^{i-1} \big[ I(Y_1^{j}; S_i, Z_i | Y_1^{j-1}, Y_2^{j-1}, S_{i+1}^{n}, Z_{i+1}^{n}) 
\\&\nonumber+ I(Y_2^{j}; S_i, Z_i | Y_1^{j-1}, Y_2^{j-1}, S_{i+1}^{n}, Z_{i+1}^{n}) \big]. \\&
= \sum_{j=1}^{i-1} I(Y_1^{j}; Z_i | Y_1^{j-1}, Y_2^{j-1}, S_{i+1}^{n}, Z_{i+1}^{n})\\&=
\sum_{i=1}^{j-1} I(Y_2^{j}; Z_i \mid Y_1^{j-1}, Y_2^{j-1}, S_{i+1}, Z_{i+1})\\&=\nonumber
\sum_{j=1}^{i-1} I(Y_1^{j}; S_i | Y_1^{j-1}, Y_2^{j-1}, S_{i,n+1}, Z_{i,n+1}) \\&+
\sum_{j=1}^{i-1} I(Y_2^{j}; S_i | Y_1^{j-1}, Y_2^{j-1}, S_{i,n+1}, Z_{i,n+1}).
\end{align}

We see that (79) and (82) split into terms of the form
$I(Y_1^{i}; Z_j | Y_1^{i-1}, Y_2^{i-1}, S_{j,n+1}, Z_{j,n+1}) + I(Y_2^{i}; Z_j | Y_1^{i-1},\\ Y_2^{i-1}, S_{j,n+1}, Z_{j,n+1}) +
I(Y_1^{i}; S_j | Y_1^{i-1}, Y_2^{i-1}, S_{j,n+1},\\ Z_{j,n+1}) + I(Y_2^{i}; S_j | Y_1^{i-1}, Y_2^{i-1}, S_{j,n+1}, Z_{j,n+1}),
$
with \(i < j\), which proves equality (77). (78) follows similarly.

\emph{Proof.} So, the proof of this lemma here is complete.

The sum rate \(R_1 + R_2\) also can be bounded as follows:
\begin{align}
&n(R_1 + R_2) \leq H(M_1, M_2 | Z^n) + n\delta
\\&\nonumber
= H(M_1, M_2 | Z^n) - H(M_1 | Y_1^n, M_2) + H(M_1 | Y_1^n, M_2)\\&\nonumber
- H(M_2 | Y_2^n, M_1) + H(M_2 | Y_2^n, M_1) + n\delta
\\&\nonumber
\leq H(M_1, M_2 | Z^n) - H(M_1 | Y_1^n, M_2)
- H(M_2 | Y_2^n, M_1)\\& + n(\epsilon_1(n) + \epsilon_2(n) + \delta)
\\&\nonumber
= H(M_1 | M_2) + H(M_2 | M_1) - H(M_1, M_2)\\&\nonumber
+ H(M_1, M_2 | Z^n) - H(M_1 | Y_1^n, M_2)
- H(M_2 | Y_2^n, M_1) \\&+ n(\epsilon_1(n) + \epsilon_2(n) + \delta)
\\&\nonumber
\leq I(M_1; Y_1^n | M_2) + I(M_2; Y_2^n | M_1)
- I(M_1, M_2; Z^n) \\&+ n(\epsilon_1(n) + \epsilon_2(n) + \delta)
\\&\nonumber
\leq I(M_1, M_2; Y_1^n) + I(M_1, M_2; Y_2^n)
- I(M_1, M_2; Z^n) + \\&\nonumber (\epsilon_1(n) + \epsilon_2(n) + \delta)
\\&\nonumber
= \sum_{i=1}^n I(M_1, M_2; Y_1^{i} | Y_1^{i-1})
+ \sum_{i=1}^n I(M_1, M_2; Y_2^{i} | Y_2^{i-1})\\&
- \sum_{i=1}^n I(M_1, M_2; Z_1^{i} | Z_i^{n+1})
+ n(\epsilon_1(n) + \epsilon_2(n) + \delta)
\\&\nonumber
= \sum_{i=1}^n I(M_1, M_2, Y_2^{i-1}, Z_i^{n+1}, S_i^{n+1}; Y_1^{i} | Y_1^{i-1})\\&\nonumber
- \sum_{i=1}^n I(Y_2^{i-1}, Z_i^{n+1}, S_i^{n+1}; Y_1^{i} | M_1, M_2, Y_1^{i-1})
\\&\nonumber
+ \sum_{i=1}^n I(M_1, M_2, Y_1^{i-1}, Z_i^{n+1}, S_i^{n+1}; Y_2^{i} | Y_2^{i-1})\\&\nonumber
- \sum_{i=1}^n I(Y_1^{i-1}, Z_i^{n+1}, S_i^{n+1}; Y_2^{i} | M_1, M_2, Y_2^{i-1})
\\&\nonumber
- \sum_{i=1}^n I(M_1, M_2, Y_1^{i-1}, Y_2^{i-1}, S_i^{n+1}; Z_i | Z_i^{n+1})
\\&\nonumber+ \sum_{i=1}^n I(Y_1^{i-1}, Y_2^{i-1}, S_i^{n+1}; Z_i | M_1, M_2, Z_i^{n+1})\\&
+ n(\epsilon_1(n) + \epsilon_2(n) + \delta)
\\&\nonumber
= \sum_{i=1}^n I(V_i, Y_{1,i})
+ \sum_{i=1}^n I(V_i, Y_{2,i})
- \sum_{i=1}^n I(V_i, Z_i)
+ n(\epsilon_1(n)  \\&+\epsilon_2(n) + \delta),
\end{align}
where the first inequality follows from the perfect secrecy condition (10). (89) is due to Fano's inequality and the last equality follows from the definition of the auxiliary random variables (42). In the above inequalities, $\epsilon_1(n)$, $\epsilon_2(n)$, and $\delta$ tend to zero as $n \to \infty$. Next, we introduce a random variable $Q$ that is independent of all other random variables and uniformly distributed over $\{1, \dots, n\}$ and define $U \triangleq (U_Q, Q)$, $V \triangleq (V_Q, Q)$, $S \triangleq (S_Q, Q)$, $Y_1 \triangleq Y_{1,Q}$, $Y_2 \triangleq Y_{2,Q}$, and $Z \triangleq Z_Q$. We obtain for the individual rate (62), (63), (10), and (14)
\begin{align}
&\nonumber R_1 \leq I(V_Q; Y_{1,Q} | Q) - \max\{ I(V_Q; S_Q | Q), I(V_Q; Z_Q | Q) \} \\&\nonumber+ (\epsilon_1(n) + \delta)
\leq I(V; Y_1) - \max\{ I(V; S), I(V; Z) \} \\&+ (\epsilon_1(n) + \delta), 
\\&\nonumber
R_2 \leq I(V_Q; Y_{2,Q} | Q) - \max\{ I(V_Q; S_Q | Q), I(V_Q; Z_Q | Q) \} \\&\nonumber+ (\epsilon_2(n) + \delta)
\leq I(V; Y_2) - \max\{ I(V; S), I(V; Z) \} +\\& (\epsilon_2(n) + \delta) 
\end{align}
and further the sum rate (64) and (95)
\begin{align}
(R_1 + R_2) 
&\leq \min \Big\{ I(V_Q; Y_{1,Q} | U_Q, Q) + I(V_Q; Y_{2,Q} | U_Q, Q) \nonumber \\
&\quad - I(V_Q; Z_Q | U_Q, Q), \; I(V_Q; Y_{1,Q} | Q) + I(V_Q; Y_{2,Q} | Q) \nonumber \\
&\quad - I(V_Q; Z_Q | Q) \Big\} + \left( \epsilon_1(n) + \epsilon_2(n) + \delta \right) \nonumber \\
&\leq \min \Big\{ I(V; Y_1 | U) + I(V; Y_2 | U) - I(V; Z | U), \nonumber \\
&\quad I(V; Y_1) + I(V; Y_2) - I(V; Z) \Big\} + \left( \epsilon_1(n) + \epsilon_2(n) + \delta \right)
\end{align}
which establishes inequalities in Theorem 2 as an outer bound on the secrecy capacity region for a discrete memoryless 2-receiver BC with non-causal CSI at the transmitter and complementary MSI at receivers. $\square$

\textbf{Theorem 3.} An achievable secrecy rate region for a discrete memoryless 2-receiver BC with non-causal CSI at the transmitter and non-complementary MSI at receivers is given by the set of all rate pairs $(R_1, R_2) \in \mathbb{R}_+^2$ that satisfy
\begin{equation}
R_1 \leq I(V; Y_2) - \max\{ I(V, S), I(V, Z) \} - R_2, 
\end{equation}
for random variables with joint probability distribution
$p(s, v, x, y_1, y_1, z) = p(s) p(v|s) p(x|v, s) p(y_1, y_1, z|x, s).$

{\bf{Proof.}} The proof of Theorem 3 is similar to Theorem 1 with negligible changes, and for simplicity, it is omitted. $\square$

\textbf{Theorem 4.} An outer bound on the secrecy capacity region for a discrete memoryless 2-receiver BC with non-causal CSI at the transmitter and non-complementary MSI at receivers is given by the set of all rate pairs $(R_1, R_2) \in \mathbb{R}_+^2$ that satisfy the following conditions:
\begin{equation}
    R_2 \leq I(V; Y_2) - \max\{ I(V, S), I(V, Z) \}, 
\end{equation}
\begin{equation}
R_1 + R_2 \leq \min \Big\{ I(V; Y_1) - \max\{ I(V; S), I(V; Z) \}, 
I(V; Y_1 | U) + I(V; Y_2 | U) - I(V; Z | U), \\
I(V; Y_1) + I(V; Y_2) - I(V; Z) \Big\}
\end{equation}
for random variables $U - V - X - (Y_1, Y_2, Z)$.

\textbf{Proof.} The proof of Theorem 4 is similar to that of Theorem 2. To show the desired outer bound on the secrecy capacity region, we need a version of Fano’s lemma suitable for the BC with receiver SI, given by
\begin{align}
&H(\mathcal{M}_1 | Y_1^n) \leq n \epsilon \left( \frac{1}{n} \right), \\& H(\mathcal{M}_2 | Y_1^n, \mathcal{M}_1) \leq n \epsilon_1(n),\\& H(\mathcal{M}_2 | Y_2^n, \mathcal{M}_1) \leq n \epsilon_2(n),
\end{align}
with $\epsilon \left( \frac{1}{n} \right), \epsilon_2(n) \to 0$ as $n \to \infty$. The derivation of $R_2 \leq I(V; Y_2) - \min \{ I(V, S), I(V, Z) \}$ is the same as that in Theorem 2. In this case, the sum rate $R_1 + R_2$ can be bounded as follows:
\begin{equation}
    n(R_1 + R_2) \leq H(M_1, M_2 | Z^n) + n \delta, 
\end{equation}
\begin{equation}
    \leq H(M_1, M_2) + n \delta,
\end{equation}
\begin{equation}
    = H(M_1, M_2 | Y_1^n) + I(M_1, M_2; Y_1^n) + n \delta, 
\end{equation}
\begin{equation}
    = H(M_1 | Y_1^n) + H(\mathcal{M}_2 | Y_1^n, \mathcal{M}_1) + I(M_1, M_2; Y_1^n) + n \delta, 
\end{equation}
\begin{equation}
    = I(M_1, M_2; Y_1^n) + n \epsilon_1(n) + n \epsilon_1(n) + n \delta, 
\end{equation}
\begin{equation}
    \leq \sum_{i=1}^{n} I(M_1, M_2; Y_1^{i} | Y_1^{i-1}) + 2n \epsilon \left( \frac{1}{n} \right) + n \delta, 
\end{equation}
\begin{equation}
    \leq \sum_{i=1}^{n} I(M_1, M_2, Y_1^{i-1}, Y_2^{i-1}, Z_i^{n-1}; Y_1^{i}) + n (2 \epsilon_1(n) + \delta), 
\end{equation}
\begin{equation}
    = \sum_{i=1}^{n} I(M_1, M_2, Y_1^{i-1}, Y_2^{i-1}, Z_i^{n-1}; Y_1^{i}) - \sum_{i=1}^{n} I(S_i^{n+1}; Y_1^{i} | M_1, M_2, Y_1^{i-1}, Y_2^{i-1}, Z_i^{n-1}) + n(2 \epsilon_1(n) + \delta), 
\end{equation}
\begin{equation}
    = \sum_{i=1}^{n} I(M_1, M_2, Y_1^{i-1}, Y_2^{i-1}, Z_i^{n-1}; Y_1^{i}) - \sum_{i=1}^{n} I(Y_1^{i-1}; S_i | M_1, M_2, S_i^{n+1}, Y_2^{i-1}, Z_i^{n-1}) + n(2 \epsilon_1(n) + \delta), 
\end{equation}
\begin{equation}
    = \sum_{i=1}^{n} I(M_1, M_2, Y_1^{i-1}, Y_2^{i-1}, Z_i^{n-1}; Y_1^{i}) - \sum_{i=1}^{n} I(M_1, M_2, S_i^{n+1}, Y_1^{i-1}, Y_2^{i-1}, Z_i^{n-1}; S_i) + n (2 \epsilon_1(n) + \delta). 
\end{equation}
\begin{equation}
    = \sum_{i=1}^{n} I(V_i; Y_1^{i}) - \sum_{i=1}^{n} I(V_i; S_i) + n (2 \epsilon_1(n) + \delta). 
\end{equation}
where the first inequality follows from the perfect secrecy condition and (110) is due to Fano’s inequality. (113) follows from the Csiszar-Korner identity. Finally, the last equality follows from the definition of the auxiliary random variables (42). Accordingly, from the perfect secrecy condition (10), the sum rate $R_1 + R_2$ can be bounded as follows:
\begin{equation}
    n(R_1 + R_2) \leq \sum_{i=1}^{n} I(V_i; Y_1^{i}) - \sum_{i=1}^{n} I(V_i; Z_i) + n (2 \epsilon_1(n) + \delta). 
\end{equation}
The derivation of another region for $R_1 + R_2$ is the same as that in the previous Theorem (Theorem 2) with negligible variations in proof. So, we have
\begin{equation}
    n(R_1 + R_2) \leq \sum_{i=1}^{n} I(V_i; Y_1^{i} | U_i) + \sum_{i=1}^{n} I(V_i; Y_2^{i} | U_i) - \sum_{i=1}^{n} I(V_i; Z_i | U_i) + n (\epsilon_1(n) + \epsilon_2(n) + \delta), 
\end{equation}
and also
\begin{equation}
    n(R_1 + R_2) \leq \sum_{i=1}^{n} I(V_i, Y_1^{i}) + \sum_{i=1}^{n} I(V_i, Y_2^{i}) - \sum_{i=1}^{n} I(V_i, Z_i) + n (\epsilon_1(n) + \epsilon_2(n) + \delta). 
\end{equation}
In the above inequalities, according to the proof of Theorem 2, $\epsilon \left( \frac{1}{n} \right)$, $\epsilon_2(n)$, and $\delta$ tend to zero as $n \to \infty$. By using a time-sharing RV $Q$ similar to the proof of Theorem 2, the inequalities in Theorem 4 are an outer bound on the secrecy capacity region for a discrete memoryless 2-receiver BC with non-causal CSI at the transmitter and non-complementary MSI at receivers. $\square$

\section{The Gaussian Versions for Inner Bound}

In this subsection, we investigate the effect of CSI at the transmitter on the achievable secrecy rate region, under the assumption of a Gaussian memoryless channel model. Specifically, we focus on deriving an inner bound on the secrecy capacity region for the considered channel configurations (referred to as Channel Model 1 and Channel Model 2).

It is well-established through Costa’s seminal work [35] that the capacity of an AWGN channel with SI or interference non-causally available at the transmitter is equal to the capacity of the corresponding channel without such interference. This fundamental result, known as "writing on dirty paper," implies that in a point-to-point Gaussian setting, the presence of non-causally known interference at the transmitter does not degrade the channel capacity.

Extending this concept, we now explore how non-causally known SI influences the secrecy performance in a multi-user wiretap scenario. To this end, we first define the system models under the Gaussian framework, including the structure of the legitimate receivers and the eavesdropper. Following that, we derive an achievable inner bound on the secrecy rate region, taking into account the presence of side information at the transmitter.

The received signals at the legitimate users and the eavesdropper are given by the following channel equations:
\begin{align}
    Y_1 &= X + S + N_1,  \\
    Y_2 &= X + S + N_2,  \\
    Z &= X + S + N_3, 
\end{align}
where $X$ represents the channel input, $S$ is the additive Gaussian SI non-causally known at the transmitter, and $N_1$, $N_2$, and $N_3$ represent the i.i.d additive Gaussian noise with zero mean and variances $\sigma_1^2$, $\sigma_2^2$, and $\sigma_3^2$ at the legitimate receivers and the eavesdropper, respectively. 

We assume that transmitted power is limited to $E[X^2] \leq P$. Since the channels are degraded, $\sigma_1^2 \leq \sigma_2^2 \leq \sigma_3^2$. Similarly to [33], we consider $V = X + \alpha S$, where $X$ and $S$ are independent random variables distributed according to $\mathcal{N}(0, P)$ and $\mathcal{N}(0, Q)$, respectively, and $\alpha$ is a parameter to be determined. 

The following theorem illustrates the Gaussian version of the achievable secrecy rate region of our channel for two cases.

{\bf{Theorem 5.}} An inner bound on the secrecy rate region of the degraded Gaussian BC with non-causal CSI at the transmitter in two cases is as follows:
\begin{enumerate}
    \item \text{Complementary MSI at Receivers}
    \begin{align}
        R_1 &= 
\begin{cases} 
    C\left(\frac{P}{\sigma_1^2}\right), & \alpha_{-1} \geq \alpha_3 \\
    & \text{or } \alpha_{-3} \leq \alpha_1 \leq \alpha_{-1}, \\
    C\left(\frac{(\sigma_3^2 - \sigma_1^2)(P + Q)}{2\sigma_1(P + Q + \sigma_3^2)}\right), & \alpha_{-3} \leq \alpha_1 \leq \alpha_3,
\end{cases}
    \end{align}
    \begin{align}
        R_2 &= 
\begin{cases} 
    C\left(\frac{P}{\sigma_2^2}\right), & \alpha_{-2} \geq \alpha_3 \\
    & \text{or } \alpha_{-3} \leq \alpha_2 \leq \alpha_{-2}, \\
    C\left(\frac{(\sigma_3^2 - \sigma_2^2)(P + Q)}{2\sigma_2(P + Q + \sigma_3^2)}\right), & \alpha_{-3} \leq \alpha_2 \leq \alpha_3.
\end{cases}
    \end{align}
    
    \item \text{Non-complementary MSI at Receivers}
    \begin{align}
       R_1 &= 
\begin{cases} 
    C\left(\frac{P}{\sigma_1^2}\right), & \alpha_{-1} \geq \alpha_3 \\
    & \text{or } \alpha_{-3} \leq \alpha_1 \leq \alpha_{-1}, \\
    C\left(\frac{(\sigma_3^2 - \sigma_1^2)(P + Q)}{2\sigma_1(P + Q + \sigma_3^2)}\right), & \alpha_{-3} \leq \alpha_1 \leq \alpha_3.
\end{cases}
    \end{align}
    \begin{align}
        R_2 &= 
        \begin{cases} 
            C\left(\frac{P}{\sigma_2^2}\right) - C\left(\frac{P}{\sigma_1^2}\right), & \alpha_{-2} \leq \alpha_3 \leq \alpha_{-1}, \\
            C\left(\frac{(\sigma_1^2 - \sigma_2^2)(P + Q)}{2\sigma_2(P + Q + \sigma_1^2)}\right), & \alpha_{-3} \leq \alpha_2 \leq \alpha_3.
        \end{cases} 
    \end{align}
\end{enumerate}
Here, $C(x) = \frac{1}{2}\log(1 + x)$, $\alpha_i = \frac{P}{P + \sigma_i^2}(1 + \sqrt{\frac{P + Q}{Q + \sigma_i^2}})$, and $\alpha_{-i} = \frac{P}{P + \sigma_i^2}(1 - \sqrt{\frac{P + Q}{Q + \sigma_i^2}})$ for $i = 1, 2, 3$.

\textbf{Proof.} For the proof of this theorem, first, we calculate the value of $I(V; Y_1)$, $I(V; Y_2)$, $I(V; S)$, and $I(V; Z)$. We have the following equality:
\[
I(V; Y_1) = H(Y_1) + H(V) - H(V, Y_1) = H(X + S + N_1)\]\[ + H(X + \alpha S) - \frac{1}{2} \log(2\pi e)^2 \left| P P + Q \alpha Q + \sigma_1^2 P P + \alpha \alpha Q^2 Q \right|
\]
\[
= \frac{1}{2} \log(2\pi e)^2 ( (P + Q + \sigma_1^2)(P + \alpha^2 Q) - \frac{1}{2} \log(2\pi e)^2 \left( (P + Q + \sigma_1^2)(P + \alpha^2 Q) - (P + \alpha Q)^2 \right) 
\tag{127}\]
\[
= \frac{1}{2} \log \frac{(P + Q + \sigma_1^2)(P + \alpha^2 Q)}{PQ(1 - \alpha)^2 + (P + \alpha^2 Q) \sigma_1^2}. \tag{128}
\]
Similarly,
\[
I(V; Y_2) = \frac{1}{2} \log \frac{(P + Q + \sigma_2^2)(P + \alpha^2 Q)}{PQ(1 - \alpha)^2 + (P + \alpha^2 Q) \sigma_2^2}, \tag{129}
\]
\[
I(V; S) = H(V) - H(V | S) = H(X + \alpha S) - H(X) = \frac{1}{2} \log \frac{(P + \alpha^2 Q)}{P}, \tag{130}
\]
\[
I(V; Z) = H(V) + H(Z) - H(V, Z) = H(X + \alpha S) +\]\[ H(X + S + N_3) - \frac{1}{2} \log(2\pi e)^2 \left| P P + \alpha \alpha Q^2 Q P P + Q \alpha Q + \sigma_3^2 \right|
= \frac{1}{2} \log \frac{(P + Q + \sigma_3^2)(P + \alpha^2 Q)}{PQ(1 - \alpha)^2 + (P + \alpha^2 Q) \sigma_3^2}. \tag{131}
\]
It is a straightforward consequence that:
\[
I(V; Y_1) - I(V; S) = \frac{1}{2} \log \frac{P(P + Q + \sigma_1^2)}{PQ(1 - \alpha)^2 + (P + \alpha^2 Q) \sigma_1^2}, \tag{132}
\]
\[
I(V; Y_1) - I(V; Z)= \frac{1}{2} \log \frac{(P + Q + \sigma_1^2)(PQ(1 - \alpha)^2 + (P + \alpha^2 Q) \sigma_3^2)}{(P + Q + \sigma_3^2)(PQ(1 - \alpha)^2 + (P + \alpha^2 Q) \sigma_1^2)}, \tag{133}
\]
\[
I(V; Y_2) - I(V; S) = \frac{1}{2} \log \frac{P(P + Q + \sigma_2^2)}{PQ(1 - \alpha)^2 + (P + \alpha^2 Q) \sigma_2^2}, \tag{134}
\]
\[
I(V; Y_2) - I(V; Z) = \frac{1}{2} \log \frac{(P + Q + \sigma_2^2)(PQ(1 - \alpha)^2 + (P + \alpha^2 Q) \sigma_3^2)}{(P + Q + \sigma_3^2)(PQ(1 - \alpha)^2 + (P + \alpha^2 Q) \sigma_2^2)}. \tag{135}
\]
Thus, we distinguish the following cases for $j = 1, 2$:
\begin{itemize}
    \item \text{Case 1:} $I(V; Y_j) \geq I(V; S) \geq I(V; Z)$,
    \item \text{Case 2:} $I(V; Y_j) \geq I(V; Z) \geq I(V; S)$.
\end{itemize}

For $j = 1$, in case 1, considering the inequality $I(V; Y_1) \geq I(V; S)$, we have:
\[
\frac{1}{2} \log \frac{(P + Q + \sigma_1^2)(P + \alpha^2 Q)}{PQ(1 - \alpha)^2 + (P + \alpha^2 Q) \sigma_1^2} \geq \frac{1}{2} \log \frac{(P + \alpha^2 Q)}{P}.
\tag{136}\]

After simplifying, under the assumption that $P, Q \geq 0$, we have:
\[
I(V; Y_1) \geq I(V; S) \iff \alpha_{-1} \leq \alpha \leq \alpha_1 \tag{137}
\]
where $\alpha_1 = \frac{P}{P + \sigma_1^2} \left( 1 + \sqrt{P + Q + \sigma_1^2} \right)$ and $\alpha_{-1} = \frac{P}{P + \sigma_1^2} \left( 1 - \sqrt{P + Q + \sigma_1^2} \right).$

Now, let us consider the inequality $I(V; S) \geq I(V; Z)$. We have:
\[
\frac{1}{2} \log \frac{(P + \alpha^2 Q)}{P} \geq \frac{1}{2} \log \frac{(P + Q + \sigma_3^2)(P + \alpha^2 Q)}{PQ(1 - \alpha)^2 + (P + \alpha^2 Q) \sigma_3^2},\tag{138}
\]
so
\[
I(V; S) \geq I(V; Z) \iff \alpha \geq \alpha_3 \text{ or } \alpha \leq \alpha_{-3} ,\tag{139}
\]
where $\alpha_3 = \frac{P}{P + \sigma_3^2} \left( 1 + \sqrt{P + Q + \sigma_3^2} \right)$ and $\alpha_{-3} = \frac{P}{P + \sigma_{23}} \left( 1 - \sqrt{P + QQ + \sigma_{32}} \right).$
In this case,
\[
I(V; Y_1) - \max \left\{ I(V, S), I(V, Z) \right\} = I(V; Y_1) - I(V, S).\tag{140}
\]
For \( j = 2 \), in case 1, considering the inequality
\[
I(V; Y_2) \geq I(V; S),\tag{141}
\]
we have
\[
\frac{1}{2} \log \left( \frac{(P + Q + \sigma_{22})(P + \alpha^2 Q)}{PQ(1 - \alpha)^2 + (P + \alpha^2 Q)\sigma_{22}} \right) \geq \frac{1}{2} \log \left( \frac{P + \alpha^2 Q}{P} \right),\tag{142}
\]
after simplifying, under the assumption that \( P, Q \geq 0 \), we have
\[
I(V; Y_2) \geq I(V; S) \iff \alpha_{-2} \leq \alpha \leq \alpha_2, \tag{143}
\]
where $\alpha_2 = \frac{P}{P + \sigma_{22}} \left( 1 + \sqrt{P + QQ + \sigma_{22}} \right)$
and
$\alpha^{-2} = \frac{P}{P + \sigma_{22}} \left( 1 - \sqrt{P + QQ + \sigma_{22}} \right).$
In this case,
\[
I(V; Y_2) - \max \left\{ I(V, S), I(V, Z) \right\} = I(V; Y_2) - I(V, S). \tag{144}
\]
In case 2, we consider the inequality \( I(V; Z) \geq I(V; S) \). Similar to above, we have
\[
I(V; Z) \geq I(V; S) \iff \alpha_{-3} \leq \alpha \leq \alpha_3, \tag{145}
\]
Thus, in this case, we have
\[
I(V; Y_1) - \max \left\{ I(V, S), I(V, Z) \right\} = I(V; Y_1) - I(V, Z), \tag{146}
\]
and
\[
I(V; Y_2) - \max \left\{ I(V, S), I(V, Z) \right\} = I(V; Y_2) - I(V, Z). \tag{147}
\]
Easy comparisons demonstrate that \( \alpha_1 \leq \alpha_2 \leq \alpha_3 \) and \( \alpha_{-3} \leq \alpha_{-2} \leq \alpha_{-1} \). Now, we investigate the properties of \( R_j \), in different cases with respect to \( \alpha \) for \( j = 1, 2 \). In case 1, for \( j = 1 \), consider
\[
R_1 = I(V; Y_1) - I(V; S), \tag{148}
\]
as defined in (123). \( R_1 \) is an increasing function with respect to \( \alpha \) as \( \alpha \leq \alpha_{\text{max}} = \frac{P}{P + \sigma_{21}} \) and is a decreasing function as \( \alpha \geq \alpha_{\text{max}} = \frac{P}{P + \sigma_{21}} \). \( R_1 \) is maximized at \( \alpha = \alpha_{\text{max}} \). In particular, we have
\[
R_1(\alpha = \alpha_{\text{max}}) = \frac{1}{2} \log \left( 1 + \frac{P}{2 \sigma_1} \right). \tag{149}
\]
In case 2, for \( j = 1 \), consider
\[
R_1 = I(V; Y_1) - I(V; Z), \tag{150}
\]
as defined in (123). \( R_1 \) is an increasing function with respect to \( \alpha \) as \( \alpha_{\text{min}} = -\frac{P}{Q} \leq \alpha \leq 1 \) and is a decreasing function as \( \alpha \leq \alpha_{\text{min}} = -\frac{P}{Q} \) or \( \alpha \geq 1 \). \( R_1 \) is minimized at \( \alpha = \alpha_{\text{min}} \) and maximized at \( \alpha = 1 \). In particular,
\[
R_1(\alpha = \alpha_{\text{min}}) = 0, \tag{151}
\]
and
\[
R_1(\alpha = 1) = \frac{1}{2} \log \left( 1 + \frac{(\sigma_{32} - \sigma_{12})(P + Q)}{2 \sigma_1 (P + Q + \sigma_{32})} \right). \tag{152}
\]
Accordingly, for \( j = 2 \) in case 1, consider
\[
R_2 = I(V; Y_2) - I(V; S), \tag{153}
\]
as defined in (124). \( R_2 \) is an increasing function with respect to \( \alpha \) as \( \alpha \leq \alpha_{\text{max}} = \frac{P}{P + \sigma_{22}} \) and is a decreasing function as \( \alpha \geq \alpha_{\text{max}} = \frac{P}{P + \sigma_{22}} \). \( R_2 \) is maximized at \( \alpha = \alpha_{\text{max}} \). In particular,
\[
R_2(\alpha = \alpha_{\text{max}}) = \frac{1}{2} \log \left( 1 + \frac{P}{2 \sigma_2} \right). \tag{154}
\]

Finally, for \( j = 2 \) in case 2, consider \( R_2 = I(V; Y_2) - I(V; Z) \) as defined in (124). \( R_2 \) is an increasing function with respect to \( \alpha \) as \( \alpha_{\text{min}} = -\frac{P}{Q} \leq \alpha \leq 1 \) and is a decreasing function as \( \alpha \leq \alpha_{\text{min}} = -\frac{P}{Q} \) or \( \alpha \geq 1 \). \( R_2 \) is minimized at \( \alpha = \alpha_{\text{min}} \) and maximized at \( \alpha = 1 \). In particular,
\[
R_2(\alpha = \alpha_{\text{min}}) = 0, \tag{155}
\]
and
\[
R_2(\alpha = 1) = \frac{1}{2} \log \left( 1 + \frac{(\sigma_3^2 - \sigma_2^2)(P + Q)}{2\sigma_2 (P + Q + \sigma_3^2)} \right). \tag{156}
\]
Thus, Theorem 5 (I) was proved. The proof of (II) in this Theorem is similar to (I) with negligible changes in (102) and for simplicity, it is omitted. \(\square\)

Now, we provide graphical illustrations to complement the theoretical findings and enhance the understanding of the achievable secrecy performance. These figures focus on analyzing the secrecy rate regions, comparing bounds, and evaluating the impact of system parameters.

Fig. 3 demonstrates the achievable secrecy rate regions for both complementary and non-complementary RMSI under varying SNR conditions. The curves illustrate the boundaries of the achievable secrecy rates for legitimate receivers, emphasizing the role of CSI in enhancing performance. As you can see, the complementary RMSI scenario shows higher achievable secrecy rates compared to the non-complementary RMSI scenario. This is because complementary RMSI reduces the decoding burden on each receiver, allowing them to achieve higher secrecy rates. Also, as SNR increases, the achievable secrecy rate increases for both RMSI configurations, which highlights the importance of SNR in improving secrecy performance. However, the complementary RMSI case shows a more significant rate increase, demonstrating the benefits of having distinct MSI available at the receivers.

\begin{figure}[ht]
		\centering
\includegraphics[width=10cm, height=7cm]{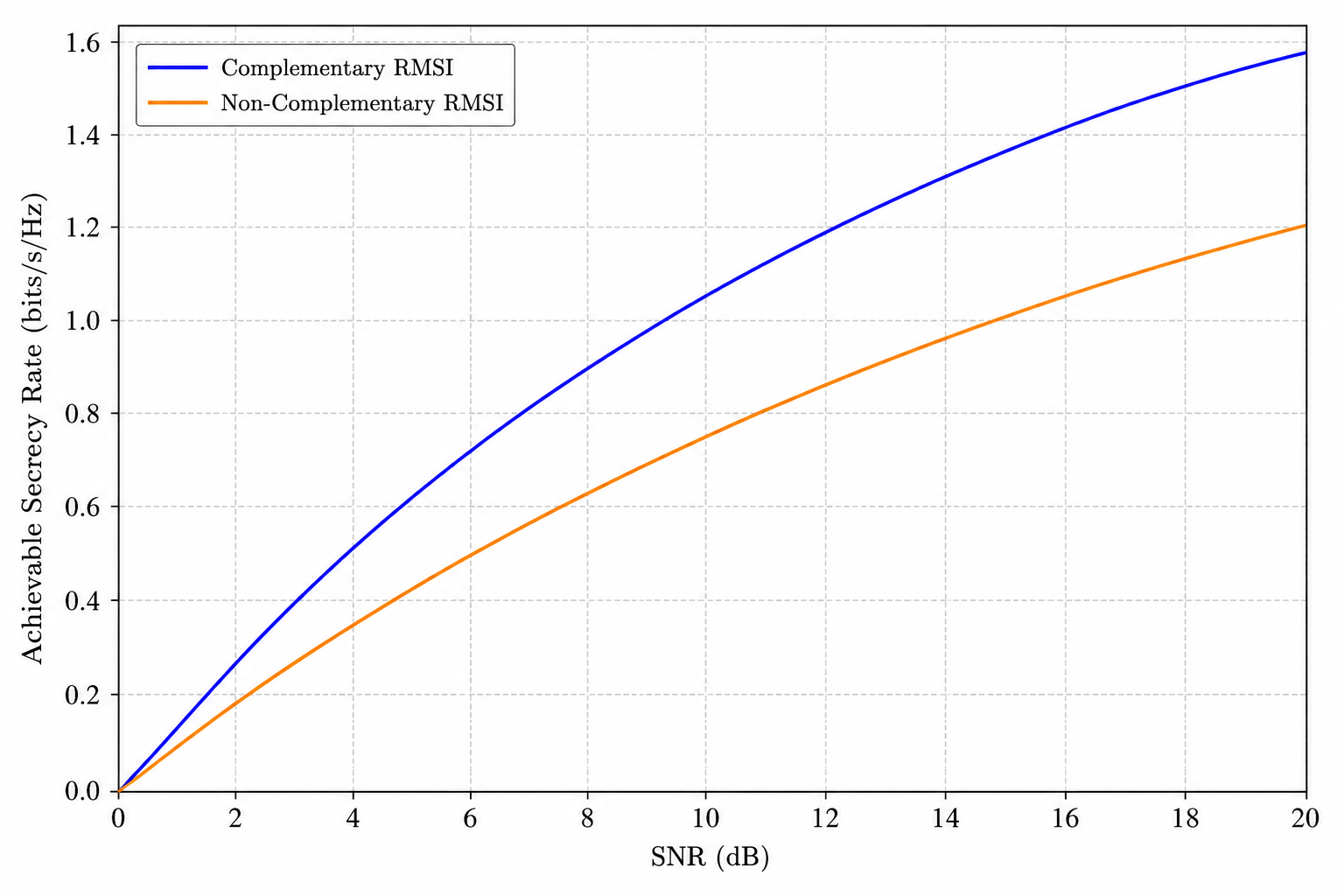}
		\caption{An illustration of the achievable secrecy rate regions.}
		\label{fig:episode}
	\end{figure}

The comparison of inner and outer bounds for secrecy capacity is shown in Fig. 4. It highlights the gap between these bounds and conditions where the inner bound approaches the outer bound, showcasing the efficiency of the proposed coding strategies. The inner bound closely approximates the outer bound, especially for higher SNR values. This suggests that the proposed coding scheme effectively utilizes the available resources, achieving secrecy rates near the theoretical maximum. Also, as you can see, at the lower SNR values, the gap between the inner and outer bounds is more noticeable. This discrepancy is attributed to the practical limitations of the coding scheme, where imperfect CSI and noise contribute to performance degradation.

\begin{figure}[ht]
		\centering
\includegraphics[width=10cm, height=7cm]{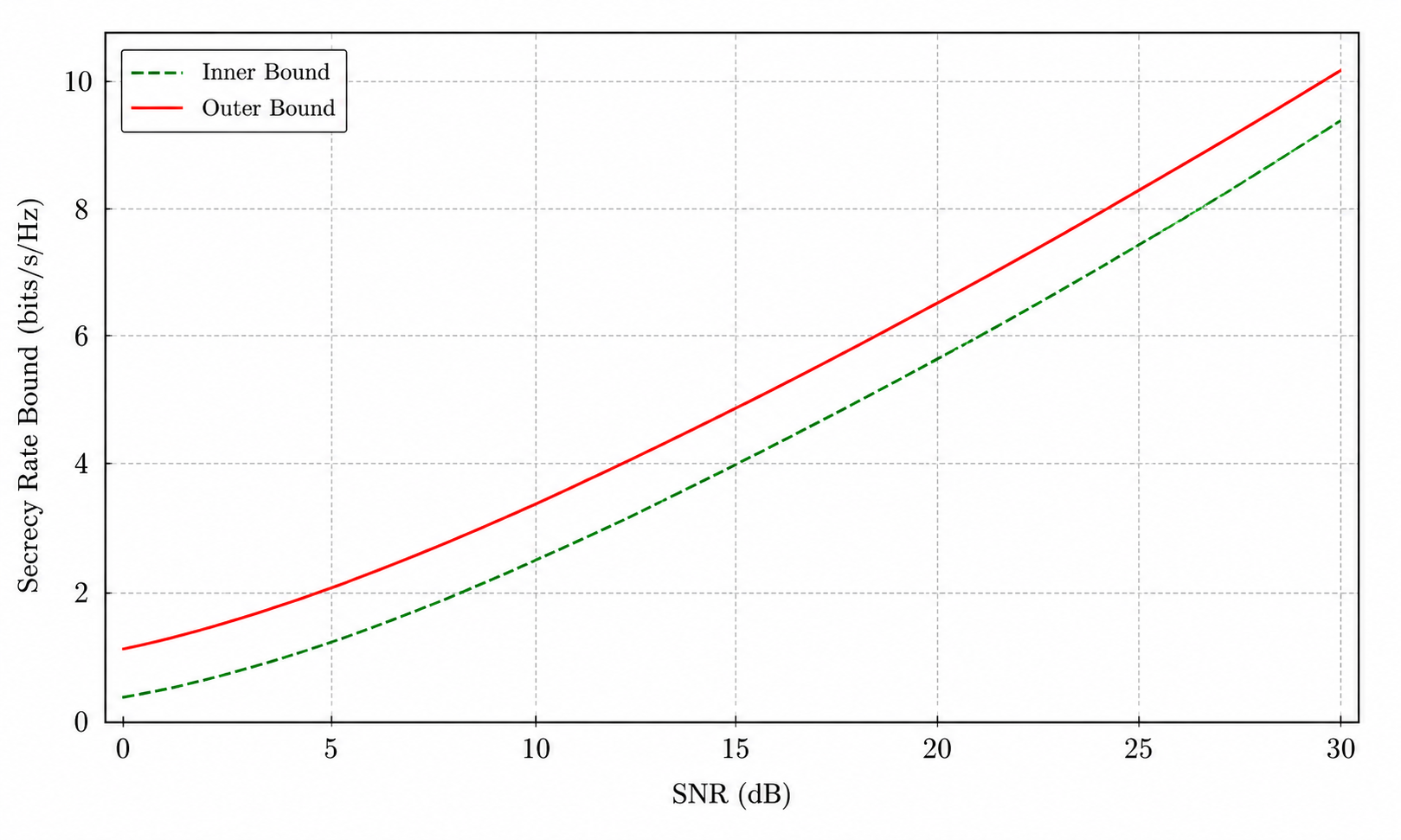}
		\caption{An illustration of the comparison of inner and outer bounds}
		\label{fig:episode2}
	\end{figure}

\section{Conclusion}


This paper has investigated the two-user degraded broadcast channel (BC) with non-causal channel state information (CSI) available at the transmitter and with receiver message side information (RMSI). We considered two distinct scenarios based on the nature of the side information: one with complementary receiver side information (RSI) and the other with non-complementary RSI. For this generalized setup, we derived an achievable secrecy rate region as well as an outer bound on the secrecy capacity region. Our proposed model extends several existing frameworks studied in the literature. Unlike earlier works on BCs with confidential messages and RSI, our analysis accommodates a more general class of receiver configurations and incorporates the availability of non-causal CSI at the transmitter. This extension reflects practical communication systems more closely, where CSI can often be estimated or acquired in advance.
Moreover, for the Gaussian instantiation of the considered broadcast channel, we demonstrated that the presence of CSI at the transmitter yields substantial improvements in the achievable secrecy rate region. These findings underscore the critical role of transmitter-side CSI in enhancing secure communication in multiuser broadcast settings and provide useful insights for the design of future secure wireless systems.


\begin{thebibliography}{35}

\bibitem{1} T. Cover, ``Broadcast channels,'' \textit{IEEE Transactions on Information Theory}, vol. 18, no. 1, pp. 2--14, 1972.
\bibitem{2} K. Marton, ``A coding theorem for the discrete memoryless broadcast channel,'' \textit{IEEE Transactions on Information Theory}, vol. 25, no. 3, pp. 306--311, 1979.
\bibitem{3} A. E. Gamal, ``The capacity of a class of broadcast channels,'' \textit{IEEE Transactions on Information Theory}, vol. 25, no. 2, pp. 166--169, 1979.
\bibitem{4} P. Bergmans, ``Random coding theorem for broadcast channels with degraded components,'' \textit{IEEE Transactions on Information Theory}, vol. 19, no. 2, pp. 197--207, 1973.
\bibitem{5} R. G. Gallager, ``Capacity and coding for degraded broadcast channels,'' \textit{Problemy Peredachi Informatsii}, vol. 10, no. 3, pp. 3--14, 1974.
\bibitem{ahlswede1975} R. Ahlswede and J. Körner, ``Source encoding with side information and a converse for degraded broadcast channels,'' \textit{IEEE Transactions on Information Theory}, vol. 6, 1975.
\bibitem{korner1977} J. Körner and K. Marton, ``General broadcast channels with degraded message sets,'' \textit{IEEE Transactions on Information Theory}, vol. 23, no. 1, pp. 60--64, 1977.
\bibitem{csiszar1978} I. Csiszár and J. Körner, ``Broadcast channels with confidential messages,'' \textit{IEEE Transactions on Information Theory}, vol. 24, no. 3, pp. 339--348, 1978.
\bibitem{shannon1958} C. E. Shannon, ``Channels with side information at the transmitter,'' \textit{IBM Journal of Research and Development}, vol. 2, no. 4, pp. 289--293, 1958.
\bibitem{10} S. Pakravan and G. A. Hodtani, ``Secrecy Capacity Outer Bound of Broadcast Channels with States Known at the Transmitter and Message Side Information at Receivers,'' in \textit{28th Iranian Conference on Electrical Engineering (ICEE)}, Tabriz, Iran, pp. 1-5, 2020.
\bibitem{11} S. Pakravan and G. A. Hodtani, ``Semi-deterministic Broadcast Channel With Side Information: A Secrecy Capacity Outer Bound,'' in \textit{10th International Conference on Computer and Knowledge Engineering (ICCKE)}, Mashhad, Iran, pp. 245-249, 2020.
\bibitem{12} S. Pakravan and G. A. Hodtani, ``An extension of Cover–Chiang work on point to point channel to less noisy broadcast channel and analysis of the receivers cognition impacts,'' in \textit{Telecommunication Systems Journal}, vol. 84, no. 4, pp. 507–519, 2023.



\bibitem{gelfand1980} S. I. Gelfand, ``Coding for channel with random parameters,'' \textit{Problems of Control and Information Theory}, vol. 9, no. 1, pp. 19--31, 1980.
\bibitem{cover2002} T. M. Cover and M. Chiang, ``Duality between channel capacity and rate distortion with two-sided state information,'' \textit{IEEE Transactions on Information Theory}, vol. 48, no. 6, pp. 1629--1638, 2002.
\bibitem{anzabi2012} N. S. Anzabi-Nezhad, G. A. Hodtani, and M. M. Kakhki, ``Information theoretic exemplification of the receiver recognition and a more general version for the Costa theorem,'' \textit{IEEE Communications Letters}, vol. 17, no. 1, pp. 107--110, 2012.
\bibitem{anzabi2015} N. S. Anzabi-Nezhad, G. A. Hodtani, and M. M. Kakhki, ``A new and more general capacity theorem for the Gaussian channel with two-sided input-noise dependent state information,'' \textit{arXiv preprint arXiv:1507.04924}, 2015.
\bibitem{rosenzweig2005} A. Rosenzweig, Y. Steinberg, and S. Shamai, ``On channels with partial channel state information at the transmitter,'' \textit{IEEE Transactions on Information Theory}, vol. 51, no. 5, pp. 1817--1830, 2005.
\bibitem{steinberg2005} Y. Steinberg, ``Coding for the degraded broadcast channel with random parameters, with causal and noncausal side information,'' \textit{IEEE Transactions on Information Theory}, vol. 51, no. 8, pp. 2867--2877, 2005.
\bibitem{steinberg2005b} Y. Steinberg and S. Shamai, ``Achievable rates for the broadcast channel with states known at the transmitter,'' in \textit{Information Theory, 2005. ISIT 2005. Proceedings. International Symposium on}, pp. 2184--2188, IEEE, 2005.
\bibitem{khosravi2011} R. Khosravi-Farsani and F. Marvasti, ``Capacity bounds for multiuser channels with non-causal channel state information at the transmitters,'' in \textit{Information Theory Workshop (ITW), 2011 IEEE}, pp. 195--199, IEEE, 2011.
\bibitem{shannon1949} C. E. Shannon, ``Communication theory of secrecy systems,'' \textit{Bell System Technical Journal}, vol. 28, no. 4, pp. 656--715, 1949.
\bibitem{wyner1975} A. D. Wyner, ``The wire-tap channel,'' \textit{Bell System Technical Journal}, vol. 54, no. 8, pp. 1355--1387, 1975.
\bibitem{liu2008} R. Liu, I. Maric, P. Spasojevic, and R. D. Yates, ``Discrete memoryless interference and broadcast channels with confidential messages: Secrecy rate regions,'' \textit{IEEE Transactions on Information Theory}, vol. 54, no. 6, pp. 2493--2507, 2008.
\bibitem{bagherikaram2008} G. Bagherikaram, A. S. Motahari, and A. K. Khandani, ``Secure broadcasting: The secrecy rate region,'' in \textit{Communication, Control, and Computing, 2008 46th Annual Allerton Conference on}, pp. 834--841, IEEE, 2008.
\bibitem{salehkalaibar2013} S. Salehkalaibar, M. Mirmohseni, and M. R. Aref, ``One-receiver two-eavesdropper broadcast channel with degraded message sets,'' \textit{IEEE Transactions on Information Forensics and Security}, vol. 8, no. 7, pp. 1162--1172, 2013.
\bibitem{chia2012} Y.-K. Chia and A. El Gamal, ``Three-receiver broadcast channels with common and confidential messages,'' \textit{IEEE Transactions on Information Theory}, vol. 58, no. 5, pp. 2748--2765, 2012.
\bibitem{mansour2014} A. S. Mansour, R. F. Schaefer, and H. Boche, ``Secrecy measures for broadcast channels with receiver side information: Joint vs individual,'' in \textit{Information Theory Workshop (ITW), 2014 IEEE}, pp. 426--430, IEEE, 2014.
\bibitem{wyrembelski2011} R. F. Wyrembelski, A. Sezgin, and H. Boche, ``Secrecy in broadcast channels with receiver side information,'' in \textit{Asilomar Conference on Signals, Systems, and Computers}, pp. 290--294, 2011.
\bibitem{chen2014} Y. Chen, O. O. Koyluoglu, and A. Sezgin, ``On the achievable individual-secrecy rate region for broadcast channels with receiver side information,'' in \textit{ISIT}, pp. 26--30, 2014.
\bibitem{chen2017} Y. Chen, O. O. Koyluoglu, and A. Sezgin, ``Individual secrecy for the broadcast channel,'' \textit{IEEE Transactions on Information Theory}, vol. 63, no. 9, pp. 5981--5999, 2017.
\bibitem{bloch2013} M. R. Bloch and J. N. Laneman, ``Strong secrecy from channel resolvability,'' \textit{IEEE Transactions on Information Theory}, vol. 59, no. 12, pp. 8077--8098, 2013.
\bibitem{tan2018} J. Y. Tan, L. Ong, and B. Asadi, ``The secure two-receiver broadcast channel with one-sided receiver side information,'' in \textit{2018 IEEE Information Theory Workshop (ITW)}, pp. 1--5, IEEE, 2018.
\bibitem{hou2014} J. Hou and G. Kramer, ``Effective secrecy: Reliability, confusion and stealth,'' in \textit{Information Theory (ISIT), 2014 IEEE International Symposium on}, pp. 601--605, IEEE, 2014.



\bibitem{saeid1} Pakravan S, Hodtani GA. ``An extension of Cover–Chiang work on point to point channel to less noisy broadcast channel and analysis of the receivers cognition impacts,'' \textit{Telecommunication Systems}, vol. 84, no. 4, pp. 507--519, 2023.



\bibitem{saeid2} Pakravan S, Hodtani GA. ``Capacity region for wireless more capable broadcast channel with channel state available at the receivers,'' \textit{in 28th Iranian conference on electrical engineering (ICEE)}, IEEE, pp. 1--5, 2020.


\bibitem{33} B. Asadi, L. Ong and S. J. Johnson. ``A unified inner bound for the two-receiver memoryless broadcast channel with channel state and message side information,'' \textit{arXiv, https://arxiv.org/abs/1601.03121}, Jan. 2016.



\bibitem{34} B. Asadi, L. Ong and S. J. Johnson. ``A unified inner bound for the two-receiver memoryless broadcast channel with channel state and message side information,'' \textit{in IEEE International Symposium on Information Theory (ISIT)}, Barcelona, Spain, pp. 175--179, 2016.




\bibitem{costa1983} M. Costa, ``Writing on dirty paper,'' \textit{IEEE Transactions on Information Theory}, vol. 29, no. 3, pp. 439--441, May 1983.
\end{thebibliography}

\end{document}